\documentclass[seceq]{ptptex}
\newcommand{\e}{\mathrm{e}}
\newcommand{\vev}[1]{\left\langle #1 \right\rangle}
\newcommand{\Sf}{S_{\mathrm{free}}}
\newcommand{\Si}{S_{\mathrm{int}}}
\newcommand{\bS}{\bar{S}}
\newcommand{\bSi}{\bar{S}_{\mathrm{int}}}
\newcommand{\hSi}{\hat{S}_{\mathrm{int}}}
\newcommand{\Sz}{S^{(0)}}
\newcommand{\Siz}{\Si^{(0)}}

\newcommand{\Sione}{\Si^{(1)}}
\newcommand{\K}[1]{K\left( #1/\Lambda \right)}
\newcommand{\D}[1]{\Delta\left( #1/\Lambda \right)}
\newcommand{\Kb}[1]{K_b \left( #1/\Lambda \right)}
\newcommand{\Kf}[1]{K_f \left( #1/\Lambda \right)}
\newcommand{\Db}[1]{\Delta_b \left( #1/\Lambda \right)}
\newcommand{\Df}[1]{\Delta_f \left( #1/\Lambda \right)}
\newcommand{\fmslash}[1]{\hbox{$#1$\kern-0.5em\raise0.3ex\hbox{/}}}
\newcommand{\Ld}[1]{\frac{\overrightarrow{\delta}}{\delta #1}}
\newcommand{\Rd}[1]{\frac{\overleftarrow{\delta}}{\delta #1}}

\newcommand{\Op}{\mathcal{O}}
\newcommand{\ep}{\epsilon}
\newcommand{\nn}{\nonumber}
\newcommand{\lb}{\left\lbrace}
\newcommand{\rb}{\right\rbrace}

\newcommand{\Tr}{\mathrm{Tr}\,}

\title{
Construction of a Wilson action for the Wess-Zumino model
}

\author{
Hidenori \textsc{Sonoda}${}^1$ %
and Kayhan \textsc{\"Ulker}${}^2$ %
}

\inst{
${}^1$ Physics Department, Kobe University, Japan\\
${}^2$ Feza G\"ursey Institute, Istanbul, Turkey
}

\abst{
    We construct a Wilson action for the Wess-Zumino model by applying
    the exact renormalization group perturbatively.  Using neither
    superfields nor auxiliary fields, we construct a supersymmetric
    action only with complex scalar and Majorana spinor fields.  We
    adopt the BRST (antifield) formalism to show the consistency of
    the construction to all orders in loop expansions.  The resulting
    action has a quadratically divergent scalar mass term which is
    absent in the superfield formalism.
}

\begin{document}

\maketitle

\section{Introduction\label{intro}}

The purpose of this paper is to construct a Wilson action for the
Wess-Zumino model by applying the exact renormalization group (ERG)
perturbatively.  ERG was originally introduced to define the continuum
limit of a quantum field theory non-perturbatively.\cite{wk} A field
theory is defined by an action with a momentum cutoff $\Lambda$.  The
renormalization group (RG) transformation is the change of the action
as we lower $\Lambda$, while we keep the physics of the theory intact.
This transformation generates an infinite number of interaction terms.
The $\Lambda$ dependence of their coefficients is determined by a
functional differential equation, which we call the ERG differential
equation.  (E for exact is added to RG as a reminder that the actions
have an infinite number of terms.)  The continuum limit corresponds to
a solution of the differential equation that reaches a fixed point as
we raise $\Lambda$ to infinity.  In practice, non-perturbative
application of ERG is difficult without any approximation; its main
role has been to give us both an assurance to and an insight into the
standard construction of a continuum limit by fine-tuning a small
number of relevant parameters of a bare action.

In this paper we use ERG only perturbatively by introducing a cutoff
through the propagators.\cite{pol} Within perturbation theory it
becomes straightforward to construct the continuum limit (or
renormalized theory) directly by solving the ERG differential
equation; we obtain the continuum limit by finding a solution that
reduces to an action with a finite number of interaction terms as we
raise $\Lambda$ to infinity.  The renormalized parameters of the theory
characterize this asymptotic behavior of the solution.\cite{integral}

The presence of an ultraviolet cutoff often gives us a false
impression that it is incompatible with local symmetry.  The
perturbative construction of gauge theories in the ERG formalism was
pioneered by Becchi \cite{becchi} and Ellwanger \cite{El} among
others.  (See \cite{RW} for the background field method, and \cite{MR}
for a manifestly gauge invariant method.)  Gauge symmetry (or BRST
invariance) is realized as the invariance of the action under an
infinitesimal field transformation where the jacobian is properly
taken into account.  Hence, the theory has the full gauge invariance
even though it must be enforced order by order in loop expansions.

Here, we wish to construct the Wess-Zumino model \cite{wz}, the
simplest four-dimensional supersymmetric theory, by using ERG.  On
introducing supersymmetry, we choose to use no auxiliary fields for
two reasons.  First, auxiliary fields, necessary to close the
supersymmetry algebra off-shell, are not available to all
supersymmetric theories; we wish to build a formalism that does not
rely on the existence of auxiliary fields.  Second, the use of
auxiliary fields (superfields to be more precise) does not bring any
insight into the realization of supersymmetry in a Wilson action.
With auxiliary fields, supersymmetry is a linear symmetry, and it is
realized automatically once we adopt superfields.\cite{bovi} ERG then
becomes merely a method of regularization.  The proof of
non-renormalization of the F-term, for example, depends on the
familiar technicalities of supergraphs.\cite{gsr} We wish to construct
the model using only its supersymmetry but no extra ingredient
provided by the superfields.

In applying ERG to the Wess-Zumino model, we have an advantage that
the model is defined strictly in four dimensions.  Hence, there is no
subtlety in defining $\gamma_5$ or Majorana fermions, in contrast to
when we use dimensional regularization.  The method of dimensional
reduction can handle $\gamma_5$, but as far as we know its consistency
has not been fully demonstrated.  (For a recent review of the
dimensional reduction, see \cite{dimred} for example.)

We first introduce a most general renormalized theory
with the same field contents as the Wess-Zumino model.  This general
theory has more number of renormalized parameters than the Wess-Zumino
model.  We then reduce the number of independent parameters by
imposing supersymmetry.  Using only component fields, the
supersymmetry transformation is non-linear in fields, just like the
gauge transformation of Yang-Mills theories.  It is this non-linearity
that calls for our attention and care.

The paper is organized as follows.  In sect.~2, we give a brief
summary of the ERG formalism, taking the four dimensional $\phi^4$
theory as an example.  In sect.~3, we apply the ERG formalism to
construct the Wess-Zumino model, using only a complex scalar field and
a Majorana spinor field (equivalently, a pair of right- and left-hand
spinors).  We choose two different cutoff functions for the scalar and
spinor, since their equality is not required by supersymmetry.  We
also hope that this choice mimics some aspect of lattice realization
of supersymmetry, if the realization is possible at all.  We show that
the theory has nine parameters if only renormalizability is imposed.
We then derive an equation that gives the invariance of the action
under the supersymmetry transformation.  In sect.~4, we construct the
action up to 1-loop.  In sect.~5, we attempt to prove, to all orders
in loop expansions, that we can realize supersymmetry by fine-tuning
the parameters.  The proof fails, however, and we explain why.  To
overcome this failure, we introduce antifields that generate the
supersymmetry transformation in sect.~6.  With the antifields,
supersymmetry is reformulated as BRST invariance of the action.  The
antifields, which are classical external sources, transform under the
supersymmetry transformation, and they play the role of auxiliary
fields in making the BRST transformation nilpotent.  (We explain this
in some details for the classical action in Appendix B.)  Outside the
context of ERG, this procedure is well known.  The antifield or BRST
formalism has been introduced to the Wess-Zumino model in \cite{ps,
  hlw}, and to super Yang-Mills theories with matter in \cite{white};
the formalism has been extended also to include general global
symmetries in \cite{bhw}.  In sect.~7, we complete the perturbative
proof that we can realize supersymmetry by fine-tuning the
parameters. In sect.~8, we give brief comments on the quadratic
divergences and holomorphy.\cite{seiberg} We conclude the paper in
sect.~9.  Three appendices are given.

Before closing this introduction, we call the reader's attention that
we work in the four dimensional euclidean space throughout the paper.
Contrary to the Minkowski space, the right- and left-hand spinors that
constitute a Majorana spinor are not complex conjugate to each other.
We summarize relevant properties of two-component spinors in Appendix
A.

\section{Realization of symmetry in the ERG approach\label{review}}

To make the paper self-contained, we give a brief summary of the ERG
formalism.  For more details, we refer the reader to lecture notes
such as \cite{becchi} and \cite{gursey} and references therein.  In
this section we only consider a real scalar field $\phi$ for
simplicity.

Let $S (\Lambda)$ be a Wilson action with the momentum cutoff
$\Lambda$.  The action is given as the sum
\begin{equation}
S (\Lambda) = \Sf (\Lambda) + \Si (\Lambda)
\end{equation}
where the free action is defined by
\begin{equation}
    \Sf (\Lambda) \equiv - \int_p \frac{1}{\K{p}} \frac{1}{2} \phi
    (-p) (p^2 + m^2) \phi (p) \quad \left( \int_p \equiv \int
        \frac{d^4 p}{(2 \pi)^4} \right)
\end{equation}
The cutoff function $K (x)$ is a positive function of $x^2$, and
has the following properties:
\begin{equation}
K(x) \lb \begin{array}{l@{\quad}l}
= 1 & (x^2 < 1)\\
\rightarrow 0 & (x^2 \to \infty)
\end{array}\right.
\end{equation}
The choice of $K (x)$ is arbitrary as long as it damps sufficiently
fast ($1/x^6$ is more than enough) as $x^2 \to \infty$.

The cutoff dependence of the interaction action $\Si (\Lambda)$ is
determined by the ERG differential equation \cite{pol}
\begin{equation}
- \Lambda \frac{\partial}{\partial \Lambda} \Si = \int_p 
\frac{\D{p}}{p^2 + m^2} \frac{1}{2} \lb \frac{\delta \Si}{\delta \phi
  (-p)} \frac{\delta \Si}{\delta \phi (p)} + \frac{\delta^2
  \Si}{\delta \phi (-p) \delta \phi (p)} \rb \label{diffSi}
\end{equation}
where
\begin{equation}
\D{p} \equiv \Lambda \frac{\partial}{\partial \Lambda} \K{p}
\end{equation}
is non-vanishing only for $p^2 > \Lambda^2$.  Alternatively the ERG
differential equation can be given for the full action as
\begin{eqnarray}
- \Lambda \frac{\partial}{\partial \Lambda} S &=& \int_p
\frac{\D{p}}{p^2 + m^2} \left[ \frac{p^2 + m^2}{\K{p}} \frac{\delta
  S}{\delta \phi (p)} \phi (p)\right.\nn\\
&& \left.\qquad + \frac{1}{2} \lb \frac{\delta S}{\delta \phi (-p)}
    \frac{\delta S}{\delta \phi (p)} + \frac{\delta^2 S}{\delta \phi
      (p) \delta \phi (-p)} \rb \right] \label{diffS}
\end{eqnarray}
The ERG differential equation (\ref{diffSi}) or (\ref{diffS}) implies
the cutoff independence of the following connected correlation
functions:
\begin{equation}
\lb\begin{array}{c@{~\equiv~}l}
\vev{\phi (p) \phi (-p)}_\infty & \frac{1 - 1/\K{p}}{p^2 + m^2} +
\frac{1}{\K{p}^2} \vev{\phi (p) \phi (-p)}_{S(\Lambda)}\\
\vev{\phi (p_1) \cdots \phi (p_n)}_\infty & \prod_{i=1}^n \frac{1}{\K{p_i}}
\cdot \vev{\phi (p_1) \cdots \phi (p_n)}_{S (\Lambda)}
\end{array}\right.
\end{equation}

Note that for (\ref{diffSi}) or (\ref{diffS}) to have a unique
solution, we must constrain the asymptotic behavior of $\Si (\Lambda)$
for $\Lambda$, large compared with $m$ and the momenta carried by the
fields.  The asymptotic behavior of a renormalized theory has the form
\begin{equation}
    \Si (\Lambda) \stackrel{\Lambda \to \infty}{\longrightarrow} \int
    d^4 x \,\left[ ( \Lambda^2 a_2 + m^2 b_2 ) \frac{1}{2} \phi^2
        + c_2 \frac{1}{2} \left( \partial_\mu \phi \right)^2 + a_4
        \frac{1}{4!} \phi^4 \right]
\end{equation}
where $a_2, b_2, c_2, a_4$ depend on $\ln \Lambda/\mu$ ($\mu$ is a
renormalization scale).  The values of $b_2, c_2, a_4$ at $\Lambda =
\mu$ can be chosen arbitrarily, and the simplest choice is
\begin{equation}
b_2 (0) = c_2 (0) = 0,\quad a_4 (0) = - \lambda
\end{equation}
where $\lambda$ is a coupling constant.\cite{integral}

Let $\Phi (p)$ be a composite operator with momentum $p$.  A composite
operator is a functional of $\phi$ that satisfies the same ERG
differential equation as an infinitesimal deformation of the
interaction action.  Hence,
\begin{equation}
- \Lambda \frac{\partial}{\partial \Lambda} \Phi (p) = \mathcal{D}
\cdot \Phi (p) \label{diffcomp}
\end{equation}
where
\begin{equation}
\mathcal{D} \equiv \int_p \frac{\D{p}}{p^2 + m^2} \lb \frac{\delta
  \Si}{\delta \phi (-p)} \frac{\delta}{\delta \phi (p)} + \frac{1}{2}
\frac{\delta^2}{\delta \phi (-p) \delta \phi (p)} \rb
\end{equation}
The differential equation implies the cutoff independence of the
correlation functions
\begin{equation}
\vev{\Phi (p) \phi (p_1) \cdots \phi (p_n)}_\infty \equiv
\prod_{i=1}^n \frac{1}{\K{p_i}} \cdot \vev{\Phi (p) \phi (p_1) \cdots
  \phi (p_n)}_{S (\Lambda)} 
\end{equation}
for $n \ge 1$.  For example, the elementary field $\phi (p)$ is not a
composite operator, but
\begin{equation}
[\phi] (p) \equiv \phi (p) + \frac{1 - \K{p}}{p^2 + m^2} \frac{\delta
  \Si}{\delta \phi (-p)}\label{compphi}
\end{equation}
is one with the correlation functions
\begin{equation}
\vev{[\phi] (p) \phi (p_1) \cdots \phi (p_n)}_\infty = \vev{\phi (p)
  \phi (p_1) \cdots \phi (p_n)}_\infty\quad (n \ge 1)
\end{equation}
The composite operator $\left[\frac{1}{2} \phi^2\right] (p)$ is less
easy to define.  For non-vanishing momentum $p$, it has the following
asymptotic behavior:
\begin{equation}
    \left[\frac{1}{2} \phi^2\right] (p) \stackrel{\Lambda \to
      \infty}{\longrightarrow} z (\ln \Lambda/\mu) \,\frac{1}{2} \int_q
    \phi (p-q) \phi (q)  
\end{equation}
The $\Lambda$ dependence of the coefficient is determined by the ERG,
but the value $z (0)$ can be chosen arbitrarily; $z(0)=1$ is the
simplest choice.

Now, given an arbitrary composite operator $\Phi (p)$ of momentum $p$,
\begin{equation}
\Sigma \equiv \int_p \K{p} \left( \frac{\delta S}{\delta \phi (p)}
    \Phi (p) + \frac{\delta \Phi (p)}{\delta \phi (p)} \right)
\label{Sigmaphi}
\end{equation}
is a composite operator of zero momentum.  Let us assume that 
$\Sigma$ vanishes for some $\Lambda$.  Then 
\begin{equation}
\Sigma (\Lambda) = 0 \label{masterphi}
\end{equation}
for any $\Lambda$, since $\Sigma$ satisfies the linear equation
(\ref{diffcomp}).  (\ref{masterphi}) is
equivalent to the Ward identity
\begin{equation}
\sum_{i=1}^n \vev{ \phi (p_1) \cdots \Phi (p_i) \cdots \phi (p_n)}_\infty = 0
\end{equation}
and it implies the invariance of the functional measure
\begin{equation}
[d\phi] \,\e^{S (\Lambda)}
\end{equation}
under the infinitesimal change of field
\begin{equation}
\phi (p) \longrightarrow \phi (p) + \ep
\K{p} \Phi (p) 
\end{equation}
The second term of (\ref{Sigmaphi}) takes into account the jacobian of
this transformation.  (\ref{masterphi}) is the generic expression of
continuous symmetry in the ERG formalism, and in sect.~\ref{without}
we introduce supersymmetry in this form.  (See \cite{qed} for an
application to QED.)

Before we review the BRST formalism, let us briefly discuss how to
introduce an external source to the action.  Let $J (-p)$ be a
classical external source coupled to a composite operator $\Phi (p)$.
The action $\bS$ in the presence of $J$ satisfies the same ERG
differential equation (\ref{diffS}) as $S$.  If $\Phi$ is the
composite operator (\ref{compphi}), $\bS$ satisfies the simple
asymptotic condition
\begin{equation}
\bS (\Lambda) - S(\Lambda) \stackrel{\Lambda \to
  \infty}{\longrightarrow} \int_p J(-p) \phi (p)
\end{equation}
and the solution of both the ERG differential equation and the above
asymptotic condition is given by
\begin{equation}
    \quad\bS (\Lambda) = \Sf (\Lambda) + \int_p J(-p) \phi (p) + \frac{1}{2}
    \int_p J(-p) \frac{1 - \K{p}}{p^2 + m^2} J(p)
    + \Si (\Lambda) \left[\phi_{\mathrm{sh}}\right] \label{Jdep}
\end{equation}
where the last term is obtained from $\Si [\phi]$ by substituting the shifted
field
\begin{equation}
\phi_{\mathrm{sh}} (p) \equiv \phi (p) + \frac{1 - \K{p}}{p^2 + m^2} J(p)
\end{equation}
into the elementary field $\phi (p)$.  We can easily check
\begin{equation}
\frac{\delta \bS}{\delta J (-p)}\Big|_{J=0} = \left[ \phi \right] (p)
\end{equation}
Now, if $\Phi$ is a more complicated composite operator such as
$\left[ \frac{1}{2} \phi^2 \right]$, the asymptotic behavior has the
form
\begin{eqnarray}
\bS (\Lambda) - S (\Lambda) &\stackrel{\Lambda \to
  \infty}{\longrightarrow}& w_1 (\ln \Lambda/\mu) \int_{p,q} J(-p)
\frac{1}{2} \phi (p-q) \phi (q)\nn\\
&&\quad  + \frac{1}{2} w_2 (\ln \Lambda/\mu) \int_p J(p) J(-p)
\end{eqnarray}
We can choose $w_1(0)$ and $w_2(0)$ at will, but we cannot give a
closed formula like (\ref{Jdep}) in this case.

We now summarize the BRST formalism, an alternative way to realize
symmetry using ERG.  We introduce an antifield $\phi^*$ which has the
statistics (fermionic in this case) opposite to that of the conjugate
field $\phi$.  We introduce $\phi^* (-p)$ as a classical external
source coupled to $\Phi (p)$ of (\ref{Sigmaphi}).  We denote the
resulting action as $\bS (\Lambda)$.  Its interaction part
\begin{equation}
\bSi (\Lambda) \equiv \bS (\Lambda) - \Sf (\Lambda)
\end{equation}
satisfies the same ERG differential equation as $\Si$:
\begin{equation}
- \Lambda \frac{\partial}{\partial \Lambda} \bSi = \int_p
\frac{\D{p}}{p^2 + m^2} \frac{1}{2} \lb \frac{\delta \bSi}{\delta \phi
  (-p)} \frac{\delta \bSi}{\delta \phi (p)} + \frac{\delta^2
  \bSi}{\delta \phi (-p) \delta \phi (p)} \rb
\end{equation}
Note $\phi^*$ plays no role in ERG.  Replacing $\Phi (p)$ of
(\ref{Sigmaphi}) by the left-derivative of the action with respect to
$\phi^*$, we extend (\ref{masterphi}) to 
\begin{equation}
\bar{\Sigma} (\Lambda) = 0
\end{equation}
where $\bar{\Sigma}$ is a composite operator defined by
\begin{equation}
\bar{\Sigma} (\Lambda) \equiv \int_p \K{p} \left( \frac{\delta \bS}{\delta \phi
      (p)} \cdot \Ld{\phi^* (-p)} \bS + \frac{\delta}{\delta \phi (p)}
    \Ld{\phi^* (-p)} \bS \right) \label{bSigmaphi}
\end{equation}
The above identity, or the vanishing of $\bar{\Sigma}$, is the BRST
invariance of the action in the ERG formalism.\cite{becchi,gursey} For
any $\bS$ we can define BRST transformation by
\begin{eqnarray}
\delta_Q &\equiv& \int_p \K{p} \Bigg( \Ld{\phi^*
  (-p)}\frac{\delta}{\delta \phi (p)}\nn\\
&&\qquad\qquad     + \Ld{\phi^* (-p)} \bS \cdot
    \frac{\delta}{\delta \phi (p)} + \frac{\delta \bS}{\delta \phi
      (p)}\cdot \Ld{\phi^* (-p)} \Bigg)
\end{eqnarray}
so that, given a composite operator $\Op$, $\delta_Q \Op$ gives
another composite operator.  We then obtain
\begin{equation}
\delta_Q \bar{\Sigma} = 0
\end{equation}
for any $\bar{\Sigma}$ given by (\ref{bSigmaphi}) even if
$\bar{\Sigma} \ne 0$.  This equation provides an essential algebraic
structure that is missing in the formalism without antifields. We will
take advantage of the algebraic structure to construct a
supersymmetric Wilson action to all orders in loop expansions.  Once
we construct a BRST invariant action, satisfying $\bar{\Sigma} = 0$,
the BRST transformation automatically satisfies nilpotency
\begin{equation}
\delta_Q \delta_Q = 0
\end{equation}
but it plays no role in this paper.

\section{Construction without antifields\label{without}}

In this section we discuss how to construct a Wilson action of the
Wess-Zumino model perturbatively without using antifields.  The starting
point is the classical action:
\begin{eqnarray}
\hspace{-0.5cm}    S_{cl} &\equiv& - \int d^4 x \Bigg[ \, \bar{\chi}_L \sigma
    \cdot \partial \chi_R + \frac{1}{2} \left( m \bar{\chi}_R \chi_R +
        \bar{m} \bar{\chi}_L \chi_L \right) + \partial_\mu
    \bar{\phi} \partial_\mu \phi + |m|^2 \bar{\phi} \phi \nn\\ 
    && \quad + g \phi \frac{1}{2} \bar{\chi}_R \chi_R + 
    \bar{g} \bar{\phi} \frac{1}{2} \bar{\chi}_L \chi_L
    +  m \phi \frac{\bar{g}}{2} \bar{\phi}^2 + 
    \bar{m} \bar{\phi} \frac{g}{2} \phi^2 
    + \frac{|g|^2}{4} |\phi|^4 \, \Bigg] \label{classical}
\end{eqnarray}
The bar above a complex scalar $\phi$ or the mass parameter $m$
denotes complex conjugation, but the bar above a spinor denotes a
transpose.  (See Appendix A for our notation on two-component spinors.)
We save ${}^*$ as a superscript to denote an antifield.

In order to classify all possible interaction terms, we introduce two
types of charges.  Though they are not conserved, we can use them to
figure out which terms are allowed and which forbidden.\cite{seiberg}
\begin{enumerate}
\item $\alpha$ charge --- We assign charge $\alpha$ to $\phi$, and $-
    \alpha$ to $\bar{\phi}$.  The action is invariant under
\begin{equation}
\phi \to \e^{i \alpha} \phi,\quad \bar{\phi} \to \e^{- i \alpha}
\bar{\phi} \label{alpha}
\end{equation}
if we assign charge $- \alpha$ to $g$ and $\alpha$ to $\bar{g}$.
In other words, the phase change of the coupling
\begin{equation}
g \to \e^{- i \alpha} g,\quad \bar{g} \to \e^{i \alpha} \bar{g}
\end{equation}
does not affect physics, since this can be compensated by
(\ref{alpha}). 
\item $\beta$ charge --- We assign charge $\beta$ to $\chi_R$, and $-
    \beta$ to $\chi_L$.  The action is invariant under
\begin{equation}
\chi_R \to \e^{i \beta} \chi_R,\quad \chi_L \to \e^{- i \beta}
\chi_L \label{beta} 
\end{equation}
if we assign charge $- 2 \beta$ to $g, m$, and $2 \beta$ to
$\bar{g}, \bar{m}$.  In other words, the phase change of the
parameters
\begin{equation}
g \to \e^{- 2i \beta} g,\quad m \to \e^{- 2i \beta} m,\quad \bar{g}
\to \e^{2i \beta} \bar{g},\quad \bar{m} \to \e^{2 i \beta} \bar{m}
\end{equation}
does not affect physics, since this can be compensated by
(\ref{beta}). 
\end{enumerate}
We will impose the above properties on the Wilson action.  Hence, the
terms such as
\begin{equation}
\bar{m}^2 g^2 \phi^2,\quad |m|^2 \,\bar{m} g \phi
\end{equation}
can be generated, though they are absent in the classical action.  The
$\alpha$ and $\beta$ charges are obviously related to the standard
R-symmetry; the R-charge is reproduced by choosing
\begin{equation}
\alpha + 2 \beta = 0
\end{equation}
In the following we treat $\alpha$ and $\beta$ as independent.  (One
of the two U(1) charges of \cite{seiberg} has $\alpha = \beta = 1$,
and the other $\alpha = 1, \beta = - \frac{1}{2}$.)  The classical
action is invariant under the following supersymmetry transformation:
\begin{equation}
\lb\begin{array}{c@{~\equiv~}l}
\delta_{cl} \phi & \bar{\xi}_R  \chi_R + \eta_\mu \partial_\mu \phi
\\ \delta_{cl} \bar{\phi} 
& \bar{\xi}_L \chi_L + \eta_\mu \partial_\mu \bar{\phi}
\\ \delta_{cl} \chi_R 
& \bar{\sigma}_\mu \xi_L \partial_\mu \phi -
\left( \bar{m} \bar{\phi} + \bar{g} \,
\frac{{\bar{\phi}}^2}{2} \right) \xi_R +
\eta_\mu \partial_\mu \chi_R\\ \delta_{cl}
\chi_L & \sigma_\mu \xi_R \partial_\mu \bar{\phi}
 - \left(  m \phi + g \,
\frac{\phi^2}{2} \right) \xi_L + \eta_\mu \partial_\mu \chi_L
\end{array}\right.\label{deltacl}
\end{equation}
where $\xi_{R,L}$ are anticommuting constant spinors, and $\eta_\mu$
is an an infinitesimal constant vector.  Hence, we obtain
\begin{eqnarray}
\delta_{cl} S_{cl} &\equiv& \int_p \Bigg[ \delta_{cl} \phi (p)
    \frac{\delta S_{cl}}{\delta \phi (p)} + \delta_{cl} \bar{\phi} (p)
    \frac{\delta S_{cl}}{\delta \bar{\phi} (p)}\nn\\
&&\quad + S_{cl} \Rd{\chi_R
      (p)} \delta_{cl} \chi_R (p) + S_{cl} \Rd{\chi_L (p)} \delta_{cl}
    \chi_L (p) \Bigg] = 0 \label{mastercl}
\end{eqnarray}
We note two things here:
\begin{enumerate}
\item In (\ref{deltacl}), $\eta_\mu$ generates translation, and the
    invariance of the action under translation holds trivially.  But
    it becomes crucial to include the translation as part of the
    transformation when we introduce a BRST invariant classical action
    $\bar{S}_{cl}$ in sect.~\ref{BRST}.  As is explained in Appendix
    B, its BRST invariance is equivalent to the nilpotency of
    (generalized) $\delta_{cl}$.
\item We can give mass dimensions and $\alpha$, $\beta$ charges to the
    parameters $\xi_{R,L}, \eta_\mu$ of the supersymmetry
    transformation (\ref{deltacl}) such that the field and its
    transformation have the same dimensions and charges.  For later
    convenience, we summarize the mass dimensions and $\alpha, \beta$
    charges of fields and constants (or parameters) in a table:
\begin{center}
\begin{tabular}{ccc|ccc}
\hline
fields& dimensions& $\alpha$,$\beta$ charges& constants&
dimensions& $\alpha$,$\beta$ charges\\ 
\hline
$\phi$& $1$& $\alpha$& $g$& $0$ & $-\alpha-2\beta$\\
$\bar{\phi}$& $1$& $- \alpha$& $\bar{g}$& $0$& $\alpha + 2 \beta$\\
$\chi_R$ & $3/2$& $\beta$& $m$& $1$& $- 2 \beta$\\
$\chi_L$ & $3/2$& $- \beta$& $\bar{m}$& $1$&  $2 \beta$\\
&&&$\xi_R$& $-1/2$& $\alpha - \beta$\\
&&&$\xi_L$& $-1/2$& $- \alpha + \beta$\\
&&& $\eta_\mu$& $-1$& $0$\\
\hline
\end{tabular}
\end{center}
\end{enumerate}
\vspace{0.2cm}
To construct a Wilson action of the Wess-Zumino model, we first split
the action into the free and interaction parts:
\[
S( \Lambda) = \Sf (\Lambda) + \Si (\Lambda)
\]
where the free part is given by
\begin{eqnarray}
&&\Sf (\Lambda) \equiv - \int_p \Bigg[\,
\frac{1}{\Kb{p}} \bar{\phi} (-p) (p^2 + |m|^2) \phi (p) \nn\\
&&\quad +
\frac{1}{\Kf{p}} \left( \bar{\chi}_L (-p) i p \cdot \sigma
\chi_R (p) + \frac{m}{2} \bar{\chi}_R \chi_R + \frac{\bar{m}}{2}
\bar{\chi}_L \chi_L \right) \,\Bigg]
\end{eqnarray}
We have chosen different cutoff functions for the scalars and spinors;
as we shall see, supersymmetry does not require
\begin{equation}
K_f = K_b
\end{equation}
  The
propagators are given by
\begin{equation}
    \begin{array}{r@{~=~}l}
        \vev{\phi (p) \bar{\phi} (-p)}_{\Sf} & \Kb {p}/(p^2 + |m|^2)\\
        \vev{\chi_R (p) \bar{\chi}_L (-p)}_{\Sf} & \left(- i p \cdot
            \bar{\sigma}\right) \Kf {p} /(p^2 + |m|^2) 
        \\
        \vev{\chi_R (p) \bar{\chi}_R (-p)}_{\Sf} & \bar{m} \Kf {p}/(p^2
        + |m|^2)\\
        \vev{\chi_L (p) \bar{\chi}_L (-p)}_{\Sf} & m \Kf {p}/(p^2 +
        |m|^2)
\end{array}
\end{equation}

The interaction action $\Si$ satisfies the ERG differential
equation
\begin{eqnarray}
    &&- \Lambda \frac{\partial}{\partial \Lambda} \Si = \int_p
    \frac{\Delta_b (p/\Lambda)}{p^2 + |m|^2} \lb \frac{\delta
      \Si}{\delta \phi (p)} \cdot \frac{\delta
      \Si}{\delta \bar{\phi} (-p)} +
    \frac{\delta^2 \Si}{\delta \phi (p) \bar{\phi} (-p)} \rb\nn\\
    &&\qquad + \int_p \frac{\Delta_f (p/\Lambda)}{p^2 + |m|^2}\nn\\
    &&\, \times \left[\, \frac{\bar{m}}{2} \lb \Si \Rd{\chi_R (p)} \cdot
        \Ld{\bar{\chi}_R 
          (-p)} \Si^{(0)} - \Tr \Ld{\bar{\chi}_R
          (-p)} \Si^{(0)} \Rd{\chi_R (p)}\rb \right.\nn\\
    &&\quad + \frac{m}{2} \lb\Si^{(0)} \Rd{\chi_L (p)} \cdot \Ld{\bar{\chi}_L
          (-p)} \Si^{(0)} - \Tr \Ld{\bar{\chi}_L
          (-p)} \Si^{(0)} \Rd{\chi_L (p)}\rb \label{ERG}\\
    &&\,\, \left. + \Si \Rd{\chi_R (p)} ( - i p \cdot
        \bar{\sigma})\Ld{\chi_L (-p)} 
    \Si - \Tr ( - i p \cdot \bar{\sigma}) \Ld{\chi_L (-p)}
    \Si \Rd{\chi_R (p)}\, \right]\nn
\end{eqnarray}
where
\begin{equation}
\Db{p} \equiv \Lambda \frac{\partial}{\partial \Lambda} \Kb{p},\quad
\Df{p} \equiv \Lambda \frac{\partial}{\partial \Lambda} \Kf{p}
\end{equation}
The minus signs in front of the traces are due to the Fermi
statistics.  Note that the $\alpha$, $\beta$ charges are preserved
under the ERG transformation.

We specify a renormalized theory by the asymptotic behavior of $\Si
(\Lambda)$ for $\Lambda$ large compared with the momenta of fields and
the mass parameters.  The asymptotic behavior is parametrized as
follows:
\begin{eqnarray}
&&\Si (\Lambda) \stackrel{\Lambda \to \infty}{\longrightarrow}
\int d^4 x\, \Bigg[ z_1 (\ln \Lambda/\mu) \,\bar{\chi}_L \sigma_\mu
\partial_\mu \chi_R\nn\\
&&\quad + z_2 (\ln \Lambda/\mu) \left( \frac{m}{2}
    \bar{\chi}_R \chi_R + 
    \frac{\bar{m}}{2} \bar{\chi}_L \chi_L \right)\nn\\
&& \quad + z_3 (\ln \Lambda/\mu)\,
\partial_\mu \bar{\phi} \partial_\mu \phi
+ \lb\Lambda^2 a_4 (\ln \Lambda/\mu) + |m|^2 z_4 (\ln \Lambda/\mu) 
\rb |\phi|^2 \nn\\
&&\quad + \lb -1+z_5 (\ln \Lambda/\mu)\rb \left( g \phi
    \frac{1}{2} \bar{\chi}_R \chi_R + 
    \bar{g} \bar{\phi} \frac{1}{2} \bar{\chi}_L \chi_L \right)\nn\\
&& \quad + \lb -1+z_6 (\ln \Lambda/\mu) \rb \left( m \phi
    \frac{\bar{g}}{2} \bar{\phi}^2 + 
    \bar{m} \bar{\phi} \frac{g}{2} \phi^2 \right)\nn\\
&& \quad + \lb -1+ z_7 (\ln \Lambda/\mu) \rb
\frac{|g|^2}{4} |\phi|^4 + z_8 (\ln \Lambda/\mu)
\left( g^2 \bar{m}^2 \frac{1}{2}\phi^2 + \bar{g}^2 m^2 
    \frac{1}{2} \bar{\phi}^2 \right) \nn\\
&& \quad+ \lb \Lambda^2 a_9 (\ln \Lambda/\mu) + |m|^2 z_9
(\ln \Lambda/\mu) \rb \left( g \bar{m} \phi + \bar{g} m
    \bar{\phi} \right) \Bigg]
\end{eqnarray}
where we have adopted coordinate space notation.  The real
coefficients $a_4, a_9$ and $z_1, \cdots, z_9$ all depend on $\ln
\Lambda/\mu$, where $\mu$ is an arbitrary renormalization scale.  The
values of $z$'s at $\Lambda = \mu$ can be chosen at will, and they
constitute part of the parameters of the theory:
\begin{equation}
z_1 (0),\cdots, z_9 (0)
\end{equation}
(A brief remark on why the coefficients are real as opposed to
complex: $a_4, z_{1,3,4,7}$ are real because the action is real, and
$a_9, z_{2,5,6,7,9}$ are real additionally because the theory is CP
invariant.  In euclidean space CP transformation is given by
\begin{equation}
\begin{array}{c@{~\rightarrow~}l@{,\quad}c@{~\rightarrow~}l}
\phi (\vec{x},x_4) & \bar{\phi} (-\vec{x},x_4)& \bar{\phi} (\vec{x},x_4) &
\phi (-\vec{x},x_4)\\
\chi_R (\vec{x},x_4) & \chi_L (- \vec{x},x_4)& \chi_L (\vec{x},x_4) &
\chi_R (- \vec{x},x_4)
\end{array}
\end{equation}
We can choose both $m$ and $g$ real by choosing the phases of the
fields appropriately.  We can then demand the theory be invariant
under the above CP.)

We now introduce a supersymmetry transformation:
\begin{equation}
\lb\begin{array}{ccl}
\delta \phi (p) & \equiv& \bar{\xi}_R  [\chi_R] (p) + \eta_\mu i p_\mu [\phi]
(p)\\ \delta \bar{\phi} 
(p)&\equiv & \bar{\xi}_L [\chi_L] (p)+ \eta_\mu i p_\mu [\bar{\phi}] (p)
\\ \delta \chi_R (p)
&\equiv & \bar{\sigma}_\mu \xi_L i p_\mu [\phi] (p) -
\left( \bar{m} [\bar{\phi}] (p) + \bar{g} \cdot 
\left[\frac{{\bar{\phi}}^2}{2} \right] (p) \right) \xi_R \\
&& \quad +
\eta_\mu i p_\mu [\chi_R] (p)\\ \delta
\chi_L (p) &\equiv & \sigma_\mu \xi_R i p_\mu [\bar{\phi}] (p) 
 - \left(  m [\phi ] (p) + g \cdot 
\left[\frac{\phi^2}{2} \right] (p) \right) \xi_L \\
&& \quad + \eta_\mu i p_\mu [\chi_L] (p)
\end{array}\right. \label{susy}
\end{equation}
Here, the fields in square brackets are composite operators:
\begin{eqnarray}
[\phi ] (p) &\equiv& \phi (p) + \frac{1-\Kb{p}}{p^2 + |m|^2}
\frac{\delta \Si}{\delta \bar{\phi} (-p)}\\
\left[\bar{\phi} \right] (p) &\equiv& \bar{\phi} (p) + \frac{1 -
  \Kb{p}}{p^2  + |m|^2} \frac{\delta \Si}{\delta \phi (-p)}\\
\left[\chi_R \right] (p) &\equiv& \chi_R (p) \nn\\
&&\, + \frac{1 - \Kf{p}}{p^2 +
  |m|^2} \left( - i p \cdot \bar{\sigma}
\frac{\overrightarrow{\delta}}{\delta \bar{\chi}_L (-p)} 
+ \bar{m} \frac{\overrightarrow{\delta}}{\delta \bar{\chi}_R (-p)} 
\right) \Si \\
\left[\chi_L\right] (p) &\equiv& \chi_L (p) \nn\\
&&\, + \frac{1 - \Kf{p}}{p^2 +
  |m|^2} \left( - i p \cdot \sigma
\frac{\overrightarrow{\delta}}{\delta \bar{\chi}_R (-p)}
+ m \frac{\overrightarrow{\delta}}{\delta \bar{\chi}_L (-p)}
\right) \Si
\end{eqnarray}
These behave trivially as $\Lambda \to \infty$:
\begin{equation}
\begin{array}{ll}
[\phi] (p) \to \phi (p),& \left[\bar{\phi}\right] (p) \to \bar{\phi}
(p)\\
\left[\chi_R\right] (p) \to \chi_R (p),& \left[\chi_L\right] (p) \to
\chi_L (p)
\end{array}
\end{equation}
The composite operators $\left[\phi^2\right] (p)$,
$\left[\bar{\phi}^2\right] (p)$ are defined by ERG differential
equations (the same as those satisfied by infinitesimal deformations
of $\Si$) and their asymptotic behaviors:
\begin{eqnarray}
\left[ \frac{\phi^2}{2}
        \right] (p) &\stackrel{\Lambda \to \infty}{\longrightarrow}&
        (1+z_{10}) \frac{\phi^2}{2} (p) + 
        z_{11} \bar{g} m \phi (p) + z_{12} \bar{g}^2 m^2 \cdot (2\pi)^4
        \delta^{(4)} (p)\\ 
        \left[ \frac{\bar{\phi}^2}{2} \right] (p) &\stackrel{\Lambda
          \to \infty}{\longrightarrow}& (1+z_{10})
        \frac{\bar{\phi}^2}{2} (p) + z_{11} g \bar{m} \bar{\phi} (p) +
        z_{12} g^2 \bar{m}^2 \cdot (2\pi)^4 \delta^{(4)} (p)
\end{eqnarray}
where $z_{10,11,12}$ are real coefficients, dependent on $\ln
\Lambda/\mu$.  The values of $z_{10,11,12}$ at $\Lambda = \mu$
\begin{equation}
z_{10} (0), z_{11} (0), z_{12} (0)
\end{equation}
should be considered as additional parameters of the theory.  Hence,
altogether, the theory has twelve real parameters
\begin{equation}
z_1 (0), \cdots, z_{12} (0)
\end{equation}

We now define a bosonic composite operator $\Sigma$ by
\begin{eqnarray}
    &&\Sigma \equiv \int_p \Kb {p}
    \left[ \delta \phi (p) \frac{\delta S}{\delta \phi
          (p)} + \frac{\delta}{\delta \phi (p)} \delta \phi (p)
        + \delta \bar{\phi} (p) \frac{\delta S}{\delta \bar{\phi} (p)}
        + \frac{\delta}{\delta \bar{\phi} (p)} \delta \bar{\phi} (p)\right]
    \nn\\
    &&\, + \int_p \Kf {p} \Bigg[
    S \Rd{\chi_R (p)} \delta \chi_R (p)
    - \Tr \delta \chi_R (p) \Rd{\chi_R (p)}\nn\\
    &&\qquad\qquad\qquad\quad + S \Rd{\chi_L (p)} \delta \chi_L (p) 
 - \Tr \delta \chi_L (p) \Rd{\chi_L (p)} \Bigg] \label{Sigma}
\end{eqnarray}
where $\delta \phi$, etc., are defined by (\ref{susy}).  Supersymmetry
of the Wilson action is the vanishing of this composite operator:
\begin{equation}
\Sigma (\Lambda) = 0 \label{master}
\end{equation}
Note that if $\Sigma$ vanishes asymptotically for large $\Lambda$, it
vanishes identically, since $\Sigma$ satisfies a linear ERG
differential equation.

Let us check (\ref{master}) at tree level.  At tree level, the
interaction action, $\Siz$, satisfies a simpler ERG differential
equation
\begin{eqnarray}
    &&- \Lambda \frac{\partial}{\partial \Lambda} \Siz = \int_p
    \frac{\Delta_b (p/\Lambda)}{p^2 + |m|^2} \frac{\delta
      \Siz}{\delta \phi (p)} \cdot \frac{\delta
      \Siz}{\delta \bar{\phi} (-p)} \nn\\
    &&\, + \int_p \frac{\Delta_f (p/\Lambda)}{p^2 + |m|^2}
    \Bigg[\, \Siz \Rd{\chi_R (p)} ( - i p \cdot
    \bar{\sigma})\Ld{\chi_L (-p)} \Siz\label{ERGtree}\\
&&\, + \frac{\bar{m}}{2} \Si \Rd{\chi_R (p)} \cdot \Ld{\bar{\chi}_R 
          (-p)} \Siz
     + \frac{m}{2}  \Siz \Rd{\chi_L (p)} \cdot \Ld{\bar{\chi}_L
          (-p)} \Siz \, \Bigg]\nn
\end{eqnarray}
We impose the asymptotic condition that $\Siz$ reduces to the
interaction part of the classical action:
\begin{equation}
\hspace{-1cm}\Siz \stackrel{\Lambda \to \infty}{\longrightarrow} - \int d^4 x
\left[ g \phi \frac{1}{2} \bar{\chi}_R \chi_R + 
    \bar{g} \bar{\phi} \frac{1}{2} \bar{\chi}_L \chi_L
    +  m \phi \frac{\bar{g}}{2} \bar{\phi}^2 + 
    \bar{m} \bar{\phi} \frac{g}{2} \phi^2 
    + \frac{|g|^2}{4} |\phi|^4 \right]
\end{equation}
Then $\Siz$ is given by the tree level Feynman diagrams where the
elementary vertices from $S_{cl}$ are connected by high momentum
propagators proportional to either $1 - K_b$ or $1 - K_f$.  At tree
level, (\ref{Sigma}) simplifies to
\begin{eqnarray}
    \Sigma^{(0)} &\equiv& \int_p \Kb {p}
    \left[ \delta^{(0)} \phi (p) \frac{\delta \Sz}{\delta \phi
          (p)} + \delta^{(0)} \bar{\phi} (p) \frac{\delta \Sz}{\delta
          \bar{\phi} (p)} \right] \nn\\
    &&\, + \int_p \Kf {p} \Bigg[
    \Sz \Rd{\chi_R (p)} \delta^{(0)} \chi_R (p) 
+ \Sz \Rd{\chi_L (p)} \delta^{(0)} \chi_L (p) \Bigg] \label{mastertree}
\end{eqnarray}
where
\begin{equation}
S^{(0)} \equiv \Sf + \Siz
\end{equation}
and $\delta^{(0)} \phi$, etc., are tree-level supersymmetry
transformations.  For $\Lambda$ large compared with $m$ and the momenta
carried by the fields, (\ref{mastertree}) reduces to
\begin{equation}
\Sigma^{(0)} (\Lambda) \stackrel{\Lambda \to \infty}{\longrightarrow}
\delta_{cl} S_{cl} = 0
\end{equation}
The last equality is due to the classical supersymmetry
(\ref{mastercl}).  Hence, $\Sigma^{(0)} (\Lambda)$ vanishes
identically for arbitrary $\Lambda$.

We wish to construct $S (\Lambda)$ that satisfies (\ref{master}) to
all orders in the number of loops.  The question is twofold:
\begin{enumerate}
\item Can we satisfy (\ref{master}) by fine-tuning the twelve parameters?
\item How many parameters are left arbitrary?
\end{enumerate}
In the next section we will answer these questions at 1-loop level.

\section{1-loop results\label{oneloop}}

Let $\Sione$ be the 1-loop correction to $\Si$.  This is obtained by
solving (\ref{ERG}) approximately.  Here, we only summarize the 1-loop
corrections to $a_{4,9} (\ln \Lambda/\mu)$ and $z_i (\ln
\Lambda/\mu)\,(i=1,\cdots,12)$:
\begin{equation}
\begin{array}{c@{~=~}l}
a_4^{(1)} (\ln \Lambda/\mu) & - \frac{|g|^2}{2} \int_p \lb -
    \Delta_b (p) + 2 \Delta_f (p) \left( 1 - K_f (p)\right) \rb/p^2\\
a_9^{(1)} (\ln \Lambda/\mu) & - \frac{1}{2} \int_p \lb \Delta_f (p)
    - \Delta_b (p) \rb/p^2 
\end{array} \label{quad}
\end{equation}
and
\begin{equation}
\begin{array}{r@{~=~}l@{\quad}r@{~=~}l}
z_1^{(1)} (\ln \Lambda/\mu) & \frac{|g|^2}{(4 \pi)^2} \ln
\Lambda/\mu + c_1&
z_2^{(1)} (\ln \Lambda/\mu) & c_2\\
z_3^{(1)} (\ln \Lambda/\mu) & \frac{|g|^2}{(4 \pi)^2} \ln \Lambda/\mu
+ c_3&
z_4^{(1)} (\ln \Lambda/\mu) & - \frac{|g|^2}{(4 \pi)^2} \ln \Lambda/\mu
+ c_4\\
z_5^{(1)}  (\ln \Lambda/\mu) & c_5&
z_6^{(1)}  (\ln \Lambda/\mu) & - \frac{|g|^2}{(4 \pi)^2} \ln \Lambda/\mu
+ c_6\\
z_7^{(1)}  (\ln \Lambda/\mu) & - \frac{|g|^2}{(4 \pi)^2} \ln \Lambda/\mu
+ c_7&
z_8^{(1)}  (\ln \Lambda/\mu) & c_8\\
z_9^{(1)}  (\ln \Lambda/\mu) & c_9&
z_{10}^{(1)} (\ln \Lambda/\mu) & \frac{|g|^2}{(4 \pi)^2} \ln
\Lambda/\mu + c_{10}\\ 
|g|^2 z_{11}^{(1)} (\ln \Lambda/\mu) & \frac{|g|^2}{(4\pi)^2} \ln
\Lambda/\mu + c_{11} |g|^2&
|g|^2 z_{12}^{(1)} (\ln \Lambda/\mu) & c_{12} |g|^2
\end{array}
\end{equation}
where 
\begin{equation}
c_i \equiv z_i^{(1)} (0)
\end{equation}
are arbitrary constants.  In calculating these, we have used the
formula
\begin{equation}
\int_p \frac{\Delta_b (p) \left(1 - K_b (p)\right)^n}{p^4} =
\int_p \frac{\Delta_f (p) \left(1 - K_f (p)\right)^n}{p^4} =
\frac{1}{(4\pi)^2} \frac{2}{n+1}
\end{equation}

Next we consider $\Sigma^{(1)}$, the 1-loop correction to $\Sigma$.  It is
given by
\begin{eqnarray}
&&\Sigma^{(1)} = \int_p \Kb{p} \Bigg[ \delta^{(0)} \phi (p) \frac{\delta
      S^{(1)}}{\delta \phi (p)} + \delta^{(1)} \phi (p) \frac{\delta
      S^{(0)}}{\delta \phi (p)} \nn\\
&&\qquad\qquad\qquad + \frac{\delta}{\delta \phi (p)} \delta^{(0)}
\phi (p) + (\phi \to \bar{\phi}) \Bigg]\nn\\ 
&& \quad + \int_p \Kf{p} \Bigg[ S^{(1)} \Rd{\chi_R (p)} \delta^{(0)}
\chi_R (p) + S^{(0)} \Rd{\chi_R (p)} \delta^{(1)} \chi_R (p) \nn\\
&&\qquad\qquad - \Tr \delta^{(0)} \chi_R (p) \Rd{\chi_R (p)}
+ (R \to L) \Bigg]
\end{eqnarray}
where $\delta^{(1)} \phi$ is the 1-loop correction to $\delta \phi$,
etc.  We only need the asymptotic behavior of $\Sigma^{(1)}$, which is
calculated as
\begin{eqnarray}
\Sigma^{(1)} &\stackrel{\Lambda \to \infty}{\longrightarrow}& \int d^4 x\,
\bar{\xi}_R \chi_R \Bigg[ t_1 \partial^2 \bar{\phi} + t_2 g |m|^2
\bar{m} + t_3 |m|^2 \bar{\phi} + t_4 g^2 \bar{m}^2 \phi\nn\\
&& \quad + t_5 m \bar{g} \frac{\bar{\phi}^2}{2} + t_6 \bar{m} g
|\phi|^2 + t_7 |g|^2 \phi \frac{\bar{\phi}^2}{2} \Bigg]\nn\\
&&\, + \int d^4 x \,\bar{\xi}_R \bar{\sigma}_\mu \chi_L \cdot \partial_\mu
\bar{\phi} \left(t_8 \bar{g} \bar{\phi} + t_9 \bar{m} \right)
 + (R \leftrightarrow L, \phi \leftrightarrow \bar{\phi})
\end{eqnarray}
where $t$'s are constants with two parts:
\begin{equation}
t_i = u_i + v_i \quad(i=1,\cdots,9)
\end{equation}
$u$'s are given by $c$'s:
\begin{equation}
\begin{array}{c@{~=~}l@{\qquad}c@{~=~}l}
u_1 & c_1 - c_3&u_2 & c_9 + c_{12} |g|^2\\
u_3 & - c_2 + c_4 + c_{11} |g|^2 &
u_4 & c_8 + c_{12} |g|^2 \\
u_5 & - c_2 + c_6 + c_{10} &u_6 & - c_5 + c_6 + c_{11} |g|^2 \\
u_7 & - c_5 + c_7 + c_{10} &u_8 & c_1 - c_5 - c_{10} \\
u_9 & c_1 - c_2 - c_{11} |g|^2
\end{array}
\end{equation}
and $v$'s are given by the 1-loop integrals:
\begin{equation}
\begin{array}{c@{~=~}l}
v_1 & - |g|^2 \int_p (K_b - K_f) \lb 4 (1- K_f) + \tilde{\Delta}_f \rb /p^4\\
v_2 & 0\\
v_3 & |g|^2 \int_p \left( K_f^2 + K_f K_b
    - K_b^2 - 3 K_f + 2 K_b \right)/p^4\\
v_4 & |g|^2 \int_p (K_b - K_f) \left( 2 - K_f - K_b \right) /p^4\\
v_5 & v_6 = |g|^2 \int_p \left( - 2 K_f^3 + 6 K_f^2 -
    2 K_b^2 - 6 K_f + 4 K_b \right)/p^4\\
v_7 & 2 |g|^2 \int_p (K_f^4 - 4 K_f^3 + 6 K_f^2
- K_b^2 - 4 K_f + 2 K_b)/p^4 \\
v_8 & v_9 = - \frac{|g|^2}{4} \int_p (K_f - K_b) \lb 2
(1- K_b) - \Delta_b \rb /p^4
\end{array}\label{1loopintegrals}
\end{equation}
We have defined
\begin{equation}
\tilde{\Delta}_f (p) \equiv - 2 p^2 \frac{d}{dp^2} \Delta_f (p)
\end{equation}
Note three points about the above results:
\begin{enumerate}
\item All the integrands of (\ref{1loopintegrals}) vanish for $p^2 <
    1$.  (We have redefined $p/\Lambda$ as $p$.)  This makes the
    integrals infrared finite.
\item The values of the integrals depend on the choice of $K_{b,f}$.
\item The coefficients satisfy one algebraic constraint:
\begin{equation}
t_5 - t_6 + t_8 - t_9 = 0 \label{constraint1}
\end{equation}
\end{enumerate}
For $\Sigma^{(1)}$ to vanish, the nine constants $t_{1,\cdots,9}$
must vanish.  (\ref{constraint1}) leaves eight independent conditions
on the twelve constants $c_{1,\cdots, 12}$.  Taking
\begin{equation}
c_1, c_2, c_5, c_9
\end{equation}
as arbitrary, we obtain the rest as
\begin{equation}
\begin{array}{r@{~=~}l}
c_3 & c_1 - v_1\\
c_4 & - c_1 + 2 c_2 + v_3 + v_5 - v_6 + v_8\\
c_6 & - c_1 + c_2 + z_5 + v_5 + v_8\\
c_7 & - c_1 + 2 c_5 + v_7 + v_8\\
c_8 & c_9 - v_2 + v_4\\
c_{10} & c_1 - c_5 - v_8\\
|g|^2 c_{11} & c_1 - c_2 - v_5 + v_6 - v_8\\
|g|^2 c_{12} & - c_9 + v_2\\
c_{13} & - c_1 + v_5 - v_6 + v_8 - v_{10}
\end{array}
\end{equation}
Thus, by fine-tuning we have obtained
\begin{equation}
\Sigma^{(1)} = 0 \label{Sigma1}
\end{equation}
We have chosen $c_1, c_2, c_5, c_9$ as arbitrary constants because
they have clear physical meanings: $c_{1,2,5}$ correspond to
normalization of fields, $m$, and $g$; $c_9$ corresponds to a finite
shift of $\phi$ proportional to $\bar{g} m$ and that of $\bar{\phi}$
proportional to $g \bar{m}$.

\section{Attempt at an all order construction without
  antifields\label{attempt}}

We wish to prove that fine-tuning makes $\Sigma$ vanish.  We outline
the inductive proof and show where it fails.  Let us first introduce a
loop expansion of the action:
\begin{equation}
S (\Lambda) = \sum_{l=0}^\infty S^{(l)} (\Lambda)
\end{equation}
Correspondingly, we expand $\Sigma$ defined by (\ref{Sigma}):
\begin{equation}
\Sigma (\Lambda) = \sum_{l=0}^\infty \Sigma^{(l)} (\Lambda)
\end{equation}
We have already shown
\begin{equation}
\Sigma^{(0)} = \Sigma^{(1)} = 0
\end{equation}
We also expand the constants $z_i (0)\,(i=1,\cdots,12)$:
\begin{equation}
z_i (0) = \sum_{l=1}^\infty c_i^{(l)}
\end{equation}
where $c_i^{(l)}$ determine the asymptotic behavior of $S^{(l)}$.  In
the previous section we have constructed $c_i^{(1)}$ (called $c_i$
there).  We also adopt the same notation for the supersymmetry
transformation of the fields:
\begin{equation}
\delta \phi (p) = \sum_{l=0}^\infty \delta^{(l)} \phi (p),\,\cdots
\end{equation}

Suppose $S^{(1)}, \cdots, S^{(l-1)}$ have been constructed by
fine-tuning $c_i^{(1)}, \cdots, c_i^{(l-1)}$ such that
\begin{equation}
\Sigma^{(0)} = \Sigma^{(1)} = \cdots = \Sigma^{(l-1)} = 0 \label{hypo}
\end{equation}
We wish to determine $c_i^{(l)}$ so that
\begin{equation}
\Sigma^{(l)} = 0
\end{equation}

First, we show that the asymptotic behavior of $\Sigma^{(l)}$ has no
$\Lambda$ dependence.  We show this using ERG.  Since $\Sigma$ is a
composite operator, $\Sigma^{(l)}$ satisfies
\begin{eqnarray}
&&- \Lambda \frac{\partial}{\partial \Lambda} \Sigma^{(l)} = \int_p
\frac{\D{p}}{p^2 + |m|^2} \left[ \frac{\delta \Si^{(0)}}{\delta \phi
      (p)} \frac{\delta}{\delta \bar{\phi} (-p)} + \frac{\delta
      \Si^{(0)}}{\delta \bar{\phi} (p)} \frac{\delta}{\delta \phi
      (-p)} \right] \Sigma^{(l)}\nn\\
&&\quad + \int_p \frac{\D{p}}{p^2 + |m|^2} 
\left[ \Si^{(0)} \Rd{\chi_R (p)} \lb - i p \cdot \bar{\sigma}
    \Ld{\bar{\chi}_L (-p)} + \bar{m} \Ld{\bar{\chi}_R (-p)}\rb
\right.\nn\\
&&\qquad \left. + \Si^{(0)} \Rd{\chi_L (p)} \lb - i p \cdot \sigma
    \Ld{\bar{\chi}_R (-p)} + m \Ld{\bar{\chi}_L (-p)} \rb \right]
\Sigma^{(l)} 
\end{eqnarray}
thanks to the induction hypothesis (\ref{hypo}).  We now take
$\Lambda$ large compared with $m$ and the momenta carried by the fields.
Since $p$ in the argument of $\Delta$ is of the same order as the
momenta carried by the fields, $\Delta$ vanishes asymptotically.
Hence, the asymptotic behavior of $\Sigma^{(l)}$ is independent of
$\Lambda$.  We can then write
\begin{eqnarray}
\Sigma^{(l)} &\stackrel{\Lambda \to \infty}{\longrightarrow}& \int d^4 x\,
\bar{\xi}_R \chi_R \Bigg[ t_1^{(l)} \partial^2 \bar{\phi} + t_2^{(l)} g |m|^2
\bar{m} + t_3^{(l)} |m|^2 \bar{\phi} + t_4^{(l)} g^2 \bar{m}^2 \phi\nn\\
&& \quad + t_5^{(l)} m \bar{g} \frac{\bar{\phi}^2}{2} + t_6^{(l)} \bar{m} g
|\phi|^2 + t_7^{(l)} |g|^2 \phi \frac{\bar{\phi}^2}{2} \Bigg]\\
&&\, + \int d^4 x \,\bar{\xi}_R \bar{\sigma}_\mu \chi_L \cdot \partial_\mu
\bar{\phi} \left(t_8^{(l)} \bar{g} \bar{\phi} + t_9^{(l)} \bar{m} \right)
 + (R \leftrightarrow L, \phi \leftrightarrow \bar{\phi})\nn
\end{eqnarray}
where $t$'s are real constants independent of $\Lambda$.  (The $t$'s
are real because the action is real and CP invariant.)  The nine
terms given above exhaust the possibilities allowed by the following:
\begin{enumerate}
\item $\Sigma$ has mass dimension $0$.
\item $\Sigma$ is a bosonic scalar.
\item $\Sigma$ has zero $\alpha, \beta$ charges.
\item $\Sigma$ is linear in either $\xi_R$ or $\xi_L$. 
\end{enumerate}
There is no term proportional to $\eta_\mu$ due to the translation
invariance.

Second, we consider the dependence of $\Sigma^{(l)}$ on the $l$-loop
constants $c_i^{(l)}$.  By expanding (\ref{Sigma}) in the number of
loops, we obtain the following expression for $\Sigma^{(l)}$:
\begin{equation}
\Sigma^{(l)} = \Sigma^{(l),1} + \Sigma^{(l),2}
\end{equation}
where 
\begin{eqnarray}
&&\Sigma^{(l),1} \equiv \int_p \Kb{p} \left[ \delta^{(0)} \phi (p)
    \frac{\delta S^{(l)}}{\delta \phi (p)} + \delta^{(l)} \phi (p)
    \frac{\delta S^{(0)}}{\delta \phi (p)} + (\phi \to \bar{\phi})
\right]\\
&&\, + \int_p \Kf{p} \left[ S^{(l)} \Rd{\chi_R (p)} \delta^{(0)}
    \chi_R (p) + S^{(0)} \Rd{\chi_R (p)} \delta^{(l)} \chi_R (p)
+ (R \to L) \right]\nn
\end{eqnarray}
and
\begin{eqnarray}
&&\Sigma^{(l),2} \equiv \int_p \Kb{p} \left[
\sum_{k=1}^{l-1} \delta^{(k)} \phi (p) \frac{\delta S^{(l-k)}}{\delta
    \phi (p)} + \frac{\delta}{\delta \phi (p)} \delta^{(l-1)} \phi (p)
+ \left(\phi \to \bar{\phi}\right) \right]\nn\\
&&\, + \int_p \Kf{p} \Bigg[ \sum_{k=1}^{l-1} S^{(l-k)} \Rd{\chi_R (p)}
    \delta^{(k)} \chi_R (p) - \Tr \delta^{(l-1)} \chi_R (p) \Rd{\chi_R
      (p)} \nn\\
&&\qquad\qquad\qquad + (R \to L) \Bigg]
\end{eqnarray}
Only $\Sigma^{(l),1}$ depends on the $l$-loop constants
$c_i^{(l)}$.  We extract only the part of its asymptotic behavior that
depends on $c^{(l)}$'s:
\begin{eqnarray}
\Sigma^{(l),1} &\stackrel{\Lambda \to \infty}{\longrightarrow}& \int d^4 x\,
\bar{\xi}_R \chi_R \Bigg[ {u}_1^{(l)} \partial^2 \bar{\phi} + u_2^{(l)} g |m|^2
\bar{m} + u_3^{(l)} |m|^2 \bar{\phi} + u_4^{(l)} g^2 \bar{m}^2 \phi\nn\\
&& \quad + u_5^{(l)} m \bar{g} \frac{\bar{\phi}^2}{2} + u_6^{(l)} \bar{m} g
|\phi|^2 + u_7^{(l)} |g|^2 \phi \frac{\bar{\phi}^2}{2} \Bigg]\\
&&\, + \int d^4 x \,\bar{\xi}_R \bar{\sigma}_\mu \chi_L \cdot \partial_\mu
\bar{\phi} \left(u_8^{(l)} \bar{g} \bar{\phi} + u_9^{(l)} \bar{m}
\right) + (R \leftrightarrow L, \phi \leftrightarrow \bar{\phi})\nn
\end{eqnarray}
where
\begin{equation}
\begin{array}{c@{~=~}l@{\qquad}c@{~=~}l}
u_1^{(l)} & c_1^{(l)} - c_3^{(l)}&
u_2^{(l)} & c_9^{(l)} + c_{12}^{(l)} |g|^2\\
u_3^{(l)} & - c_2^{(l)} + c_4^{(l)} + c_{11}^{(l)} |g|^2&
u_4^{(l)} & c_8^{(l)} + c_{12}^{(l)} |g|^2\\
u_5^{(l)} & - c_2^{(l)} + c_6^{(l)} + c_{10}^{(l)}&
u_6^{(l)} & - c_5^{(l)} + c_6^{(l)} + c_{11}^{(l)} |g|^2\\
u_7^{(l)} & - c_5^{(l)} + c_7^{(l)} + c_{10}^{(l)}&
u_8^{(l)} & c_1^{(l)} - c_5^{(l)} - c_{10}^{(l)}\\
u_9^{(l)} & c_1^{(l)} - c_2^{(l)} - c_{11}^{(l)} |g|^2
\end{array}
\end{equation}
(The calculation is a repetition of what we have already done at
1-loop.)  The $u^{(l)}$ coefficients satisfy one algebraic
constraint
\begin{equation}
u_5^{(l)} - u_6^{(l)} + u_8^{(l)} - u_9^{(l)} = 0 \label{uconstraint}
\end{equation}
We note that $u_i^{(l)}$ does not give the whole $t_i^{(l)}$, but an
additional contribution, $v_i^{(l)}$, comes from $\Sigma^{(l),2}$ so
that 
\begin{equation}
t_i^{(l)} = u_i^{(l)} + v_i^{(l)}
\end{equation}
The coefficients $v_i^{(l)}$ are independent of $c^{(l)}$'s; we can
tune only $u_i^{(l)}$ through $c^{(l)}$'s.  But the algebraic
constraint (\ref{uconstraint}) implies that we cannot satisfy
\begin{equation}
t_i^{(l)} = 0 \quad (i=1,\cdots,9)
\end{equation}
unless $t^{(l)}$'s also satisfy the same constraint:
\begin{equation}
t_5^{(l)} - t_6^{(l)} + t_8^{(l)} - t_9^{(l)} = 0  \label{tconstraint}
\end{equation}
Equivalently, $v^{(l)}$'s must satisfy
\begin{equation}
v_5^{(l)} - v_6^{(l)} + v_8^{(l)} - v_9^{(l)} = 0  \label{vconstraint}
\end{equation}
But we have no control over $v^{(l)}$'s; they are fixed by the choice
of the lower loop constants $c^{(1)}, \cdots, c^{(l-1)}$.  We do not
know how to derive (\ref{tconstraint}) or (\ref{vconstraint}).  This
is where our proof fails.  We resort to the BRST formalism for help.

\section{BRST formalism\label{BRST}}

In this section we introduce antifields that generate supersymmetry
transformation.  We do this for the sole purpose of obtaining an
algebraic structure that enables us to derive the constraint
(\ref{tconstraint}) (or equivalently (\ref{vconstraint})).  Once we
know the construction works, we can discard antifields entirely and go
ahead with constructing a Wilson action in terms of fields alone.
>From now on we take
\begin{equation}
K_b = K_f = K
\end{equation}
for simplicity, even though this equality is not required by
supersymmetry.

In the BRST formalism we regard $\xi_{R,L}$ as \textbf{constant
  bosonic ghost spinors}, and $\eta_\mu$ as a \textbf{constant
  fermionic ghost vector}.  We introduce antifields (sources) such
that we reproduce the original supersymmetry transformation at the
vanishing sources.  The antifields have the opposite statistics to the
corresponding fields: $\phi^*, \bar{\phi}^*$ are fermionic, and
$\chi^*_R, \chi^*_L$ are bosonic.  Let us summarize the properties of
antifields and ghosts in a table:
\begin{center}
\begin{tabular}{cccc}
\hline
antifields/ghosts & statistics &  dimensions & $\alpha,\beta$ charges\\
\hline
$\phi^*$& f& $3$& $- \alpha$\\ 
$\bar{\phi}^*$& f& $3$& $\alpha$\\
$\chi_R^*$& b & $5/2$& $-\beta$\\
$\chi_L^*$& b & $5/2$& $\beta$\\
$\xi_R$& b& $- 1/2$& $\alpha-\beta$\\
$\xi_L$& b& $- 1/2$& $-\alpha+\beta$ \\
 $\eta_\mu$& f& $- 1$& $0$\\
\hline
\end{tabular}
\end{center}
The ghost number is assigned as
\begin{center}
\begin{tabular}{c|c}
\hline
ghost number $1$& ghost number $-1$\\
\hline
$\xi_R, \xi_L, \eta_\mu$& $\phi^*, \bar{\phi}^*, \chi_R^*, \chi_L^*$\\
\hline
\end{tabular}
\end{center}
so that the action has zero ghost number.

At the classical level, we obtain the following action:
\begin{eqnarray}
\bS_{cl} &\equiv& S_{cl} + \int d^4 x \Bigg[
\phi^* \left( \bar{\xi}_R \chi_R + \eta_\mu \partial_\mu \phi \right)
+ \bar{\phi}^* \left( \bar{\xi}_L \chi_L + \eta_\mu \partial_\mu
    \bar{\phi} \right)\nn\\
&& \, + \bar{\chi}_R^* \left( \bar{\sigma}_\mu \xi_L \partial_\mu \phi
    - \left( \bar{m} \bar{\phi} + \bar{g} \frac{\bar{\phi}^2}{2}
    \right) \xi_R + \eta_\mu \partial_\mu \chi_R \right) \\
&& \, + \bar{\chi}_L^* \left( \sigma_\mu \xi_R \partial_\mu \bar{\phi}
    - \left( m \phi + g \frac{\phi^2}{2}
    \right) \xi_L + \eta_\mu \partial_\mu \chi_L \right) 
- \bar{\chi}_R^* \xi_R \cdot \bar{\chi}_L^* \xi_L \Bigg]\nn
\end{eqnarray}
The last term, quadratic in antifields, is necessary so that
$\bS_{cl}$ satisfies the Zinn-Justin equation
\begin{eqnarray}
\bar{\Sigma}_{cl} &\equiv& \int_p \Bigg[ \, \frac{\delta \bS_{cl}}{\delta \phi
  (p)} \Ld{\phi^* (-p)} \bS_{cl} + \frac{\delta \bS_{cl}}{\delta
  \bar{\phi} (p)} \Ld{\bar{\phi}^* (-p)} \bS_{cl}\nn\\
&&\, + \bS_{cl} \Rd{\chi_R (p)} \cdot \Ld{\bar{\chi}_R^* (-p)}
\bS_{cl} + \bS_{cl} \Rd{\chi_L (p)} \cdot \Ld{\bar{\chi}_L^* (-p)}
\bS_{cl} \Bigg] \nn\\
&&\qquad - \bar{\xi}_L \sigma_\mu \xi_R
\frac{\overrightarrow{\delta}}{\delta \eta_\mu} S_{cl} = 0 \label{ZJ}
\end{eqnarray}
This is the classical limit of the BRST invariance that we will
introduce at the end of this section.

Let $\bar{S}$ be the total action in the presence of antifields.  It
is split as
\begin{equation}
\bar{S} (\Lambda) = \Sf (\Lambda) + \bar{S}_{\mathrm{int}} (\Lambda)
\end{equation}
where $\bar{S}_{\mathrm{int}}$ satisfies the same ERG differential
equation (\ref{ERG}) as $\Si$, but their asymptotic behaviors differ
due to antifields:
\begin{eqnarray}
    &&  \bSi (\Lambda) - \Si (\Lambda) \stackrel{\Lambda \to
      \infty}{\longrightarrow} \int d^4 x\,  \Bigg[ \phi^* \left(
        \bar{\xi}_R \chi_R  + \eta_\mu \partial_\mu \phi \right) +
    \bar{\phi}^* \left( \bar{\xi}_L \chi_L  + \eta_\mu \partial_\mu
        \bar{\phi} \right)\nn\\
    && \quad + \bar{\chi}_R^* \left( \bar{\sigma}_\mu \xi_L \partial_\mu \phi
        + \eta_\mu \partial_\mu \chi_R 
    \right) + \bar{\chi}_L^* \left( \sigma_\mu \xi_R \partial_\mu \bar{\phi} 
        + \eta_\mu \partial_\mu \chi_L \right) \nn\\
    &&\quad - j_R \left( (1 + z_{10}) \bar{g} \frac{\bar{\phi}^2}{2} +
        (1 + z_{11}) \bar{m} \bar{\phi} + z_{12} g \bar{m}^2 \right)\nn\\
    &&\quad - j_L \left( (1 + z_{10}) g \frac{\phi^2}{2} +
        (1 + z_{11}) m \phi + z_{12} \bar{g} m^2 \right)\nn\\
    &&\quad + \lb -1 + z_{13} (\ln \Lambda/\mu) \rb j_R j_L 
    \Bigg] \label{bSiasymp}
\end{eqnarray}
where
\begin{equation}
j_R  \equiv \bar{\chi}_R^* \,\xi_R,\quad
j_L  \equiv \bar{\chi}_L^* \,\xi_L
\end{equation}
The $\Lambda$ dependent coefficients $z_{10,11,12}$ are the same as
those in sect.~\ref{without}.  The real coefficient $z_{13}$ is
generated from the product of $[\phi^2] (p)$ and $[\bar{\phi}^2]
(-p)$.  We note three points here:
\begin{enumerate}
\item Three parameters $z_{10} (0), z_{11} (0), z_{12} (0)$ that
    define supersymmetry transformation is now incorporated into the action
    $\bar{S}_{\mathrm{int}}$.
\item $\bar{S}_{\mathrm{int}}$ depends on one extra parameter $z_{13}
    (0)$, which is absent in the definition of $\Si$ and in the supersymmetry
    transformation at the vanishing antifields.
\item The antifields $j_R, j_L$ are similar to the auxiliary fields
    that close the off-shell supersymmetry algebra.  Unlike the
    auxiliary fields, $j_R$ and $j_L$ are external sources not to be
    integrated over.
\end{enumerate}

Most antifield dependence is introduced through linear couplings to the
elementary fields, and can be given explicitly:
\begin{eqnarray}
&&\bSi (\Lambda) = \int_p \Big[\, \phi^* (-p) \left(\bar{\xi}_R \chi_R  (p) +
    \eta_\mu i p_\mu \phi (p) \right) \nn\\
&&\qquad + \bar{\phi}^* (-p) \left( \bar{\xi}_L \chi_L (p)
    + \eta_\mu i p_\mu \bar{\phi} (p) \right)\nn\\
&&\qquad + \bar{\chi}_R^* (-p) \left( \bar{\sigma}_\mu \xi_L i p_\mu
    \phi (p) - \xi_R \bar{m} \bar{\phi} (p) + \eta_\mu i p_\mu \chi_R
    (p) \right) \nn\\
&&\qquad + \bar{\chi}_L^* (-p) \left( \sigma_\mu \xi_R i p_\mu \bar{\phi} (p)
    - \xi_L m \phi (p) + \eta_\mu i p_\mu \chi_L (p) \right) \nn\\
&&\qquad\qquad - j_R (-p) j_L (p) \,\Big]\nn\\
&& \quad + \int_p \frac{1 - \K{p}}{p^2 + |m|^2} \Big[
- \bar{\xi}_R i p \cdot \bar{\sigma} \xi_L \phi^* (-p) \bar{\phi}^*
(p)\nn\\
&&\,\, + \left( \bar{\chi}_R^* (-p) i p \cdot \bar{\sigma} \xi_L - m
    \bar{\chi}_L^* (-p) \xi_L \right) \left( - \bar{\chi}_L^* (p) i p
    \cdot \sigma \xi_R - \bar{m} \bar{\chi}_R^* (p) \xi_R \right)
\Big]\nn\\
&& \quad + \hSi \left[ \phi_{\mathrm{sh}},
    \bar{\phi}_{\mathrm{sh}}, \chi_{R,\mathrm{sh}},
    \chi_{L,\mathrm{sh}}; j_R, j_L \right] \label{bSihSi}
\end{eqnarray}
where the shifted fields are defined as follows:
\begin{equation}
\hspace{-1cm}
\lb\begin{array}{ccl}
\phi_{\mathrm{sh}} (p) &\equiv& \phi (p) \\
&& \, + \frac{1 - \K{p}}{p^2 + |m|^2} \lb -
\bar{\phi}^* (p) i p_\mu \eta_\mu - \left( \bar{\chi}_L^* (p) i p
    \cdot \sigma + \bar{m} \bar{\chi}_R^* (p) \right) \xi_R \rb\\
\bar{\phi}_{\mathrm{sh}} (p) &\equiv& \bar{\phi} (p) \\
&&\, + \frac{1 - \K{p}}{p^2 +
  |m|^2} \lb - \phi^* (p) i p_\mu \eta_\mu - \left( \bar{\chi}_R^* (p)
    i p \cdot \bar{\sigma} + m \bar{\chi}_L^* (p) \right) \xi_L \rb\\
\chi_{R, \mathrm{sh}} (p) &\equiv& \chi_R (p) + \frac{1 - \K{p}}{p^2 +
  |m|^2} \big\lbrace 
\left( - \bar{m} \chi_R^* (p) + i p \cdot \bar{\sigma} \chi_L^* (p)
\right) i p_\mu \eta_\mu \\
&&\qquad\qquad\qquad\qquad\quad + \bar{m} \phi^* (p) \xi_R - i p \cdot
\bar{\sigma} \bar{\phi}^* (p) \xi_L \big\rbrace\\
\chi_{L, \mathrm{sh}} (p) &\equiv& \chi_L (p) + \frac{1 - \K{p}}{p^2 +
  |m|^2} \big\lbrace 
\left( - m \chi_L^* (p) + i p \cdot \sigma \chi_R^* (p) \right) i
p_\mu \eta_\mu \\
&&\qquad\qquad\qquad\qquad\quad + m \bar{\phi}^* (p) \xi_L - i p \cdot
\sigma \phi^* (p) \xi_R \big\rbrace
\end{array}\right.\label{shift}
\end{equation}
$\hSi$ is obtained from $\Si$ by coupling an external source $j_R$ to
the composite operator $\left[\phi^2/2\right]$, and $j_L$ to
$\left[\bar{\phi}^2/2\right]$.  $\hSi$ satisfies the same ERG
differential equation as $\bSi$; its asymptotic behavior is determined
from (\ref{bSiasymp}) as
\begin{eqnarray}
    &&\hSi (\Lambda) - \Si (\Lambda) \stackrel{\Lambda
      \to \infty}{\longrightarrow} \int d^4 x\, \left[ 
        - j_R \lb (1 + z_{10}) \bar{g} \frac{\bar{\phi}^2}{2} +
        z_{11} \bar{m} \bar{\phi}  + z_{12} g \bar{m}^2 \rb\right.\nn\\
    &&\qquad \left. - j_L \lb (1 + z_{10}) g \frac{\phi^2}{2}  +
        z_{11} m \phi + z_{12} \bar{g} m^2 \rb
    + z_{13}  \, j_R  j_L 
    \quad \right]
\end{eqnarray}
where we have suppressed the $\ln \Lambda/\mu$ dependence of $z$'s.

We now define a fermionic composite operator by
\begin{eqnarray}
&&\bar{\Sigma} \equiv \int_p \K{p} \Bigg[ \Ld{\phi^* (-p)} \bar{S}
\cdot \frac{\delta \bar{S}}{\delta \phi (p)} + \frac{\delta}{\delta
  \phi (p)} \Ld{\phi^* (-p)} \bar{S}\nn\\
&&\qquad + \Ld{\bar{\phi}^* (-p)} \bar{S} \cdot \frac{\delta
  \bar{S}}{\delta \bar{\phi} (p)} + \frac{\delta}{\delta \bar{\phi}
  (p)} \Ld{\bar{\phi}^* (-p)} \bar{S}\nn\\
&&\quad - \Tr \lb \Ld{\bar{\chi}_R^* (-p)} \bar{S} \cdot \bar{S}
  \Rd{\chi_R (p)} + \Ld{\bar{\chi}_R^* (-p)} \bar{S}
  \Rd{\chi_R (p)} \rb\nn\\
&&\quad - \Tr \lb \Ld{\bar{\chi}_L^* (-p)} \bar{S} \cdot \bar{S}
  \Rd{\chi_L (p)} + \Ld{\bar{\chi}_L^* (-p)} \bar{S}
  \Rd{\chi_L (p)} \rb \Bigg] \nn\\
&&\qquad - \bar{\xi}_L \sigma_\mu \xi_R \cdot
\frac{\overrightarrow{\partial}}{\partial \eta_\mu} \bar{S}
\label{bSigma}
\end{eqnarray}
We wish to fine-tune the thirteen parameters
\begin{equation}
z_1 (0),\cdots, z_{13} (0)
\end{equation}
for the BRST invariance
\begin{equation}
\bar{\Sigma} = 0 \label{bmaster}
\end{equation}
At the vanishing antifields this reduces to (\ref{master}), the
supersymmetry of $S$.  The classical BRST invariance (\ref{ZJ})
implies that (\ref{bmaster}) holds at tree level.

$\bar{\Sigma}$ defined by (\ref{bSigma}) satisfies the following algebraic
constraint: 
\begin{eqnarray}
&&\delta_Q \bar{\Sigma} \equiv \int_p \K{p} \Bigg[ \nn\\
&&\quad \Ld{\phi^* (-p)} \bar{\Sigma} \cdot \frac{\delta
  \bar{S}}{\delta \phi (p)} + 
\Ld{\phi^* (-p)} \bar{S} \cdot \frac{\delta \bar{\Sigma}}{\delta \phi
  (p)} + \Ld{\phi^* (-p)} \frac{\delta}{\delta \phi (p)} \bar{\Sigma}\nn\\
&&+
\Ld{\bar{\phi}^* (-p)} \bar{\Sigma} \cdot \frac{\delta \bar{S}}{\delta
  \bar{\phi} (p)} +
\Ld{\bar{\phi}^* (-p)} \bar{S} \cdot \frac{\delta \bar{\Sigma}}{\delta
  \bar{\phi} (p)} + \Ld{\bar{\phi}^* (-p)} \frac{\delta}{\delta
  \bar{\phi} (p)} \bar{\Sigma}\nn\\ 
&& \hspace{-1cm}+ \Tr \lb \Ld{\bar{\chi}_R^* (-p)} \bar{\Sigma} \cdot \bar{S}
  \Rd{\chi_R (p)} + \Ld{\bar{\chi}_R^* (-p)} \bar{S} \cdot \bar{\Sigma}
  \Rd{\chi_R (p)} + \Ld{\bar{\chi}_R^* (-p)} \bar{\Sigma}
  \Rd{\chi_R (p)} \rb\nn\\
&&\hspace{-1cm} + \Tr \lb \Ld{\bar{\chi}_L^* (-p)} \bar{\Sigma} \cdot \bar{S}
  \Rd{\chi_L (p)} + \Ld{\bar{\chi}_L^* (-p)} \bar{S} \cdot \bar{\Sigma}
  \Rd{\chi_L (p)} + \Ld{\bar{\chi}_L^* (-p)} \bar{\Sigma}
  \Rd{\chi_L (p)} \rb \Bigg]\nn\\
&&\quad - \bar{\xi}_L \sigma_\mu \xi_R \cdot
\frac{\overrightarrow{\partial}}{\partial \eta_\mu} \bar{\Sigma}
\quad = 0 \label{bSigmaconstraint}
\end{eqnarray}
We shall see, in the next section, that this provides the algebraic
constraint (\ref{tconstraint}) that we could not derive without
antifields.

\section{All order proof\label{proof}}

The all order proof in the BRST formalism proceeds analogously to the
attempted proof in sect.~\ref{attempt}. Before we begin, we make some
preparation.  As can be seen from (\ref{bSihSi}), most of the
antifield dependence of $\bS$ is given by the shift (\ref{shift}) of
the fields.  It is then natural to express $\bar{\Sigma}$ in terms of
(\ref{shift}) by replacing each field in $\bar{\Sigma}$ by the
corresponding shifted field.  We obtain
\begin{equation}
\hspace{-0.5cm}    \bar{\Sigma}
    [\phi,\bar{\phi},\chi_R,\chi_L,\phi^*,\bar{\phi}^*,\chi_R^*,\chi_L^*] 
    = \tilde{\Sigma}
    \left[\phi_{\mathrm{sh}},\bar{\phi}_{\mathrm{sh}},\chi_{R,\mathrm{sh}},
    \chi_{L,\mathrm{sh}},\phi^*,\bar{\phi}^*,\chi_R^*,\chi_L^*
\right]  \label{bSigma-tSigma}  
\end{equation}
where $\tilde{\Sigma}$ is defined by
\begin{eqnarray}
&& \tilde{\Sigma}
\left[\phi,\bar{\phi},\chi_R,\chi_L,\phi^*,\bar{\phi}^*,\chi_R^*,\chi_L^*\right]
\nn\\   
&\equiv& \int_p \K{p} \left[ \Ld{\phi^* (-p)} \tilde{S}
\cdot \frac{\delta \tilde{S}}{\delta \phi (p)} + \Ld{\bar{\phi}^*
  (-p)} \tilde{S} \cdot \frac{\delta \tilde{S}}{\delta \bar{\phi}
  (p)}\right.\nn\\
&&\, - \tilde{S} \Rd{\chi_R (p)} \cdot \Ld{\bar{\chi}_R^* (-p)}
\tilde{S} - \Tr \Ld{\bar{\chi}_R^* (-p)} \tilde{S} \Rd{\chi_R (p)}\nn\\
&&\, \left.- \tilde{S} \Rd{\chi_L (p)} \cdot \Ld{\bar{\chi}_L^* (-p)}
\tilde{S} - \Tr \Ld{\bar{\chi}_L^* (-p)} \tilde{S} \Rd{\chi_L (p)} \right]\nn\\
&&\, - \bar{\xi}_L \sigma_\mu \xi_R \cdot
\frac{\overrightarrow{\partial}}{\partial \eta_\mu} \tilde{S}
\end{eqnarray}
and $\tilde{S}$ is defined by
\begin{eqnarray}
    &&    \tilde{S}
    \left[\phi,\bar{\phi},\chi_R,\chi_L,\phi^*,\bar{\phi}^*,\chi_R^*,
        \chi_L^*\right] \nn\\ 
    &\equiv& \Sf [\phi,\bar{\phi},\chi_R,\chi_L] + \hSi
    [\phi,\bar{\phi},\chi_R,\chi_L;j_R,j_L]\nn\\ 
    && + \int_p \frac{1}{\K{p}} \Big[ \quad\phi^* (-p) \left\lbrace i p_\mu
        \eta_\mu \phi (p) + \bar{\xi}_R \chi_R (p) \right\rbrace \nn\\
    &&\quad +
    \bar{\phi}^* (-p) \left\lbrace i p_\mu \eta_\mu \bar{\phi} (p) +
        \bar{\xi}_L \chi_L (p) \right\rbrace\nn\\
    && \quad + \bar{\chi}_R^* (-p) \left\lbrace i p_\mu \eta_\mu \chi_R (p) +
        i p \cdot \bar{\sigma} \xi_L \phi (p) - \bar{m} \xi_R \bar{\phi}
        (p) \right\rbrace\nn\\
    && \quad + \bar{\chi}_L^* (-p) \left\lbrace i p_\mu \eta_\mu \chi_L (p) +
        i p \cdot \sigma \xi_R \bar{\phi} (p) - m \xi_L \phi (p)
    \right\rbrace\nn\\
    &&\quad - j_R (-p) j_L (p) \quad\Big] \label{Stilde}
\end{eqnarray}
Thus, the identity (\ref{bmaster}) is equivalent to the identity
\begin{equation}
\tilde{\Sigma}
\left[\phi,\bar{\phi},\chi_R,\chi_L,\phi^*,\bar{\phi}^*,\chi_R^*,\chi_L^*\right]
= 0 \label{tmaster}
\end{equation}
for the unshifted fields.  This identity is simpler to consider than
(\ref{bmaster}), since unshifting makes the antifield dependence of
$\tilde{\Sigma}$ simpler than that of $\bar{\Sigma}$.  Note, however,
that for $\Lambda$ large compared with $m$ and the momenta of the
fields, $\bar{\Sigma}$ and $\tilde{\Sigma}$ are the same:
\begin{equation}
\bar{\Sigma} (\Lambda) - \tilde{\Sigma} (\Lambda) \stackrel{\Lambda
  \to \infty}{\longrightarrow} 0
\end{equation}
since the shifts in (\ref{shift}) vanish in this limit.

By definition, $\tilde{\Sigma}$ satisfies its own algebraic identity:
\begin{eqnarray}
&&\tilde{\delta}_Q \tilde{\Sigma} \equiv \int_p \K{p} \Bigg[ \nn\\
&&\quad \Ld{\phi^* (-p)} \tilde{\Sigma} \cdot \frac{\delta \tilde{S}}{\delta
   \phi (p)} + 
\Ld{\phi^* (-p)} \tilde{S} \cdot \frac{\delta \tilde{\Sigma}}{\delta \phi
  (p)} + \Ld{\phi^* (-p)} \frac{\delta}{\delta \phi (p)} \tilde{\Sigma}\nn\\
&&\, +
\Ld{\bar{\phi}^* (-p)} \tilde{\Sigma} \cdot \frac{\delta \tilde{S}}{\delta
  \bar{\phi} (p)} +
\Ld{\bar{\phi}^* (-p)} \tilde{S} \cdot \frac{\delta \tilde{\Sigma}}{\delta
  \bar{\phi} (p)} + \Ld{\bar{\phi}^* (-p)} \frac{\delta}{\delta
  \bar{\phi} (p)} \tilde{\Sigma}\nn\\ 
&&\hspace{-1cm} + \Tr \lb \Ld{\bar{\chi}_R^* (-p)} \tilde{\Sigma}
\cdot \tilde{S} 
  \Rd{\chi_R (p)} + \Ld{\bar{\chi}_R^* (-p)} \tilde{S} \cdot \tilde{\Sigma}
  \Rd{\chi_R (p)} + \Ld{\bar{\chi}_R^* (-p)} \tilde{\Sigma}
  \Rd{\chi_R (p)} \rb\nn\\
&&\hspace{-1cm} + \Tr \lb \Ld{\bar{\chi}_L^* (-p)} \tilde{\Sigma}
\cdot \tilde{S} 
  \Rd{\chi_L (p)} + \Ld{\bar{\chi}_L^* (-p)} \tilde{S} \cdot \tilde{\Sigma}
  \Rd{\chi_L (p)} + \Ld{\bar{\chi}_L^* (-p)} \tilde{\Sigma}
  \Rd{\chi_L (p)} \rb \Bigg]\nn\\
&&\qquad - \bar{\xi}_L \sigma_\mu \xi_R \cdot
\frac{\overrightarrow{\partial}}{\partial \eta_\mu} \tilde{\Sigma}
\quad = 0 \label{tSigmaconstraint}
\end{eqnarray}
In fact we can show
\begin{equation}
    \quad\delta_Q \bar{\Sigma}
    \left[\phi,\bar{\phi},\chi_R,\chi_L,\phi^*,\bar{\phi}^*,\chi_R^*,\chi_L^*
    \right] = \tilde{\delta}_Q \tilde{\Sigma}
    \left[
        \phi_{\mathrm{sh}},\bar{\phi}_{\mathrm{sh}},\chi_{R,\mathrm{sh}},
        \chi_{L,\mathrm{sh}},\phi^*,\bar{\phi}^*,\chi_R^*,\chi_L^* \right]
\end{equation}
Hence, the two identities (\ref{bSigmaconstraint}),
(\ref{tSigmaconstraint}) are really the same.

We now begin our inductive proof.  We will show that we can satisfy
(\ref{bmaster}) by fine-tuning the thirteen parameters $z_1
(0),\cdots, z_{13} (0)$.  The proof proceeds analogously to the
attempted proof in sect.~\ref{attempt}. We use the same notation for
loop expansions:
\begin{equation}
\bar{S} = \sum_{l=0}^\infty \bar{S}^{(l)},\quad 
\bar{\Sigma} = \sum_{l=1}^\infty \bar{\Sigma}^{(l)},\quad
z_i (0) = \sum_{l=1}^\infty c_i^{(l)}\,(i=1,\cdots,13)
\end{equation}
As the induction hypothesis, we assume that we have chosen $c^{(1)},
\cdots, c^{(l-1)}$ so that
\begin{equation}
\bar{\Sigma}^{(0)} = \cdots = \bar{\Sigma}^{(l-1)} = 0 \label{bhypo}
\end{equation}
We need to show that fine-tuning of $c^{(l)}$'s makes
\begin{equation}
\bar{\Sigma}^{(l)} = 0
\end{equation}

First, we show that the asymptotic behavior of $\bar{\Sigma}^{(l)}
(\Lambda)$ is independent of $\Lambda$, using ERG and (\ref{bhypo}).
We omit the proof since it is exactly the same as in
sect.~\ref{attempt}.

Second, we enumerate all possible terms in the asymptotic behavior of
$\bar{\Sigma}^{(l)}$, which is the same as that of
$\tilde{\Sigma}^{(l)}$.  The asymptotic form of $\tilde{\Sigma}^{(l)}$
is parametrized by twelve real constants $t_1^{(l)},
\cdots, t_{12}^{(l)}$ as follows:
\begin{eqnarray}
&&\tilde{\Sigma}^{(l)} \stackrel{\Lambda \to \infty}{\longrightarrow}
 \int \bar{\xi}_R \chi_R \Bigg[
t_1^{(l)} \partial^2 \bar{\phi} + t_2^{(l)} g |m|^2 \bar{m} +
t_3^{(l)} |m|^2 \bar{\phi} 
+ t_4^{(l)} g^2 \bar{m}^2 \phi \nn\\
&&\qquad  + t_5^{(l)} m \bar{g} \frac{\bar{\phi}^2}{2} +
t_6^{(l)} \bar{m} g |\phi|^2 + t_7^{(l)} |g|^2 \phi \frac{\bar{\phi}^2}{2}
+ j_L \left( t_{10}^{(l)} m + t_{11}^{(l)} g \phi \right) \Bigg]\nn\\
&&\quad + \int \bar{\chi}_L \sigma_\mu \xi_R \cdot \partial_\mu
\bar{\phi} \left( t_8^{(l)} \bar{g} \bar{\phi} + t_9^{(l)} \bar{m} \right)
 + t_{12}^{(l)} \int \bar{\chi}_L \sigma_\mu \xi_R \cdot \partial_\mu j_L\nn\\
&&\,  + \left( R \leftrightarrow L, \phi \leftrightarrow
    \bar{\phi}, m \leftrightarrow \bar{m}, g \leftrightarrow \bar{g} \right)
\end{eqnarray}
This is the most general form allowed by the following constraints:
\begin{enumerate}
\item $\tilde{\Sigma}$ has mass dimension $0$.
\item $\tilde{\Sigma}$ is a fermionic scalar.
\item $\tilde{\Sigma}$ has no $\alpha,\beta$ charges.
\item $\tilde{\Sigma}$ has ghost number $1$.
\item $\tilde{\Sigma}^{(l)}$ is independent of $\eta_\mu, \phi^*,
    \bar{\phi}^*$.
\item $\tilde{\Sigma}^{(l)}$ depends on $\chi_R^*, \chi_L^*$ only
    through $j_R, j_L$.
\end{enumerate}
We derive the last two properties in Appendix C.  As in
sect.~\ref{attempt} we now divide each $t_i^{(l)}$ into two parts:
\begin{equation}
t_i^{(l)} = u_i^{(l)} + v_i^{(l)}
\end{equation}
where $u^{(l)}$'s are linear combinations of $c^{(l)}$'s, and
$v^{(l)}$'s are determined by the action up to $(l-1)$-loop level.
$u^{(l)}$'s are obtained from the asymptotic behavior of
\begin{eqnarray}
&&\tilde{\Sigma}^{(l),1} \equiv \int_p \K{p} \left[ \Ld{\phi^* (-p)}
    \tilde{S}^{(0)} \cdot \frac{\delta \tilde{S}^{(l)}}{\delta \phi (p)} 
    + (\phi \to \bar{\phi}, \phi^* \to \bar{\phi}^*) \right.\\
&&\,\, \left.- \tilde{S}^{(l)} \Rd{\chi_R (p)} \cdot \Ld{\bar{\chi}_R^*
  (-p)} \tilde{S}^{(0)} -  \tilde{S}^{(0)} \Rd{\chi_R (p)} \cdot
   \Ld{\bar{\chi}_R^* (-p)} \tilde{S}^{(l)} + (R \to L) \right]\nn
\end{eqnarray}
as follows:
\begin{equation}
\begin{array}{c@{~=~}l@{\qquad}c@{~=~}l}
u^{(l)}_1 & c^{(l)}_1 - c^{(l)}_3&
u^{(l)}_2 & c^{(l)}_9 + c^{(l)}_{12} |g|^2\\
u^{(l)}_3 & - c^{(l)}_2 + c^{(l)}_4 + c^{(l)}_{11} |g|^2&
u^{(l)}_4 & c^{(l)}_8 + c^{(l)}_{12} |g|^2\\
u^{(l)}_5 & - c^{(l)}_2 + c^{(l)}_6 + c^{(l)}_{10}&
u^{(l)}_6 & - c^{(l)}_5 + c^{(l)}_6 + c^{(l)}_{11} |g|^2\\
u^{(l)}_7 & - c^{(l)}_5 + c^{(l)}_7 + c^{(l)}_{10}&
u^{(l)}_8 & c^{(l)}_1 - c^{(l)}_5 - c^{(l)}_{10}\\
u^{(l)}_9 & c^{(l)}_1 - c^{(l)}_2 - c^{(l)}_{11} |g|^2&
u^{(l)}_{10} & - c^{(l)}_2 - c^{(l)}_{11} |g|^2 - c^{(l)}_{13}\\
u^{(l)}_{11} & - c^{(l)}_5 - c^{(l)}_{10} - c^{(l)}_{13}&
u^{(l)}_{12} & c^{(l)}_1 + c^{(l)}_{13}
\end{array}
\end{equation}
The calculation is mostly the same as that done in
sect.~\ref{attempt}.  The $u^{(l)}$'s are not linearly independent,
but they satisfy the following three linear relations:
\begin{equation}
\lb\begin{array}{r}
u_5^{(l)} - u_6^{(l)} + u_8^{(l)} - u_9^{(l)} = 0\\
u_8^{(l)} - u_{11}^{(l)} - u_{12}^{(l)} = 0\\
u_9^{(l)} - u_{10}^{(l)} - u_{12}^{(l)} = 0
\end{array}\right. \label{urelations}
\end{equation}
The first relation is the same as that found in sect.~\ref{attempt}.

Third, we derive algebraic constraints on the $t^{(l)}$ constants.
For fine-tuning to work, $t^{(l)}$'s must satisfy the same linear
relations as (\ref{urelations}).  These come from the algebraic
identity (\ref{tSigmaconstraint}).  Since $\tilde{\Sigma}$ vanishes up
to $(l-1)$-loop, we obtain
\begin{eqnarray}
&&\hspace{-0.5cm}
\left(\tilde{\delta}_Q \tilde{\Sigma}\right)^{(l)} = \int_p \K{p}
\Bigg[ \nn\\ 
&&\quad \Ld{\phi^* (-p)} \tilde{\Sigma}^{(l)} \cdot \frac{\delta
  \tilde{S}^{(0)}}{\delta \phi (p)} + \Ld{\phi^* (-p)} \tilde{S}^{(0)}
\cdot \frac{\delta \tilde{\Sigma}^{(l)}}{\delta \phi (p)} + 
\left( \phi \to \bar{\phi}, \phi^* \to
    \bar{\phi}^* \right)\\
&& - \tilde{S}^{(0)} 
  \Rd{\chi_R (p)} \cdot \Ld{\bar{\chi}_R^* (-p)} \tilde{\Sigma}^{(l)}
 +\tilde{\Sigma}^{(l)} 
  \Rd{\chi_R (p)} \cdot \Ld{\bar{\chi}_R^* (-p)} \tilde{S}^{(0)} 
+ (R \to L) \,\,\Bigg]\nn
\end{eqnarray}
Taking the asymptotic part, we obtain
\begin{eqnarray}
&&\left(\tilde{\delta}_Q
        \tilde{\Sigma}\right)^{(l)} \stackrel{\Lambda \to
      \infty}{\longrightarrow}\nn\\
&& \,\,(t_5^{(l)} - t_6^{(l)} - t_{10}^{(l)} +  t_{11}^{(l)} ) 
    \int d^4 x \,\bar{\xi}_R \chi_R \cdot
        \bar{\xi}_L \chi_L \left( \bar{m} g \phi - m \bar{g}
        \bar{\phi} \right)\nn\\
    &&\, + \int d^4 x \, \bar{\xi}_R \chi_R \cdot \bar{\xi}_L
        \sigma_\mu \partial_\mu \chi_R  \left[ (- t_9^{(l)} +
        t_{10}^{(l)} + 
        t_{12} ^{(l)}) m + (- t_8^{(l)} + t_{11}^{(l)} + t_{12}^{(l)}
        ) g \phi \right] \nn\\ 
    &&\, +  \bar{\xi}_R \bar{\sigma}_\mu \xi_L \int d^4 x
    \Bigg[ (t_5^{(l)} 
    - t_9^{(l)} - t_6^{(l)} + t_8^{(l)}) m \bar{g} \partial_\mu \phi \cdot
    \frac{\bar{\phi}^2}{2} \nn\\
    &&\qquad + \partial_\mu \phi \cdot j_L \lb (t_{10}^{(l)} -
    t_9^{(l)} + t_{12}^{(l)}) m 
    + (t_{11}^{(l)} - t_8^{(l)} + t_{12}^{(l)} ) g \phi \rb \Bigg]\nn\\
&&\, + \left( \lb \phi, \chi_R, \rb
      \leftrightarrow \lb \bar{\phi}, \chi_L \rb \right)
\end{eqnarray}
This must vanish.  Hence, $t^{(l)}$'s satisfy the same linear
relations as (\ref{urelations}): 
\begin{equation}
\lb\begin{array}{r}
t_5^{(l)} - t_6^{(l)} + t_8^{(l)} - t_9^{(l)} = 0\\
t_8^{(l)} - t_{11}^{(l)} - t_{12}^{(l)} = 0\\
t_9^{(l)} - t_{10}^{(l)} - t_{12}^{(l)} = 0
\end{array}\right.\label{trelations}
\end{equation}
This also implies
\begin{equation}
\lb\begin{array}{r}
v_5^{(l)} - v_6^{(l)} + v_8^{(l)} - v_9^{(l)} = 0\\
v_8^{(l)} - v_{11}^{(l)} - v_{12}^{(l)} = 0\\
v_9^{(l)} - v_{10}^{(l)} - v_{12}^{(l)} = 0
\end{array}\right.
\end{equation}

Last, we fine-tune $c^{(l)}$'s to make $t^{(l)}$'s vanish.  Because
of (\ref{trelations}), we have  nine independent conditions to satisfy
using thirteen constants.  Taking
\begin{equation}
c_1^{(l)}, c_2^{(l)}, c_5^{(l)}, c_9^{(l)}
\end{equation}
as arbitrary, we obtain the rest as follows:
\begin{equation}
\begin{array}{r@{~=~}l}
c^{(l)}_3 & c^{(l)}_1 - v^{(l)}_1\\
c^{(l)}_4 & - c^{(l)}_1 + 2 c^{(l)}_2 + v^{(l)}_3 + v^{(l)}_5 -
v^{(l)}_6 + v^{(l)}_8\\ 
c^{(l)}_6 & - c^{(l)}_1 + c^{(l)}_2 + c^{(l)}_5 + v^{(l)}_5 + v^{(l)}_8\\
c^{(l)}_7 & - c^{(l)}_1 + 2 c^{(l)}_5 + v^{(l)}_7 + v^{(l)}_8\\
c^{(l)}_8 & c^{(l)}_9 - v^{(l)}_2 + v^{(l)}_4\\
c^{(l)}_{10} & c^{(l)}_1 - c^{(l)}_5 - v^{(l)}_8\\
|g|^2 c^{(l)}_{11} & c^{(l)}_1 - c^{(l)}_2 - v^{(l)}_5 + v^{(l)}_6 - v^{(l)}_8\\
|g|^2 c^{(l)}_{12} & - c^{(l)}_9 + v^{(l)}_2\\
c^{(l)}_{13} & - c^{(l)}_1 + v^{(l)}_5 - v^{(l)}_6 + v^{(l)}_8 - v^{(l)}_{10}
\end{array}
\end{equation}
This concludes the proof by induction.

The physical meaning of each arbitrary parameters is clear:
\begin{enumerate}
\item $z_1 (0)$ --- overall normalization of scalar and spinor fields;
    this is unphysical. 
\item $z_2 (0)$ --- normalization of $m$
\item $z_5 (0)$ --- normalization of $g$
\item $z_9 (0)$ --- constant shift of $\phi$ proportional to $\bar{g}
    m$, and $\bar{\phi}$ proportional to $g \bar{m}$; this is also
    unphysical.
\end{enumerate}

\section{Quadratic divergences and holomorphy\label{quadratic}}

Before concluding the paper, we make remarks on the quadratic
divergences and holomorphy of the Wilson action $S (\Lambda)$.

First on the quadratic divergences, by which we mean the two leading
terms in the asymptotic behavior of the action:
\begin{equation}
S (\Lambda) \stackrel{\Lambda \to \infty}{\longrightarrow} \int d^4 x
\, \left[ \Lambda^2 a_4 (\ln \Lambda/\mu) |\phi|^2 + \Lambda^2 a_9
    (\ln \Lambda/\mu) \left( \bar{m} g \phi + m \bar{g} \bar{\phi}
    \right) \right]
\end{equation}
In sect.~\ref{oneloop} we have found the following 1-loop results:
\begin{equation}
\lb\begin{array}{c@{~=~}l}
a_4^{(1)} & - \frac{|g|^2}{2} \int_p \lb -
    \Delta_b (p) + 2 \Delta_f (p) \left( 1 - K_f (p)\right) \rb/p^2\\
a_9^{(1)} & - \frac{1}{2} \int_p \lb \Delta_f (p)
    - \Delta_b (p) \rb/p^2 
\end{array}\right.\label{a49}
\end{equation}
We observe two points and make a comment:
\begin{enumerate}
\item The values of the constants $a_4^{(1)}, a_9^{(1)}$ depend on the
    choice of the cutoff functions $K_{b,f}$, and therefore
    non-universal.   For a general choice of $K_{b,f}$, both the
    constants are non-vanishing.
\item Though supersymmetry does not require $K_b = K_f$, this choice
    would make $a_9^{(1)}$ zero, but not $a_4^{(1)}$.  
\item If we adopt the superfields, we must choose $K_b = K_f$.  The
    F-term in the superfield formalism does not receive any quantum
    correction.\cite{gsr}  This suggests a possibility of choosing the
    parameters $z_{1,2,5,9}$ so that
\begin{equation}
a_9 = 0
\end{equation}
to all orders in loop expansions.  
\end{enumerate}

The Wilson action in terms of superfields does not have any quadratic
divergence, but it is generated by integration over the auxiliary fields.
At 1-loop, it is easy to see this explicitly.  Let us give an outline
here.  The tree level action has a vertex
\begin{equation}
- |g|^2  \bar{\phi} (p_3) \bar{F} (p_4) \frac{1 - \K{(p_1+p_2)}}{(p_1+p_2)^2 +
  |m|^2} \phi (p_1) F (p_2)
\end{equation}
where $F, \bar{F}$ are auxiliary fields.  Contracting $F$ and
$\bar{F}$ using the propagator
\begin{equation}
\K{p} \frac{p^2}{p^2 + |m|^2} 
\end{equation}
we obtain the scalar mass term (at zero momentum) as follows:
\begin{eqnarray}
&&- |g|^2 \int_p \K{p} \left( 1 - \K{p} \right) \frac{p^2}{\left( p^2 +
      |m|^2 \right)^2}\nn\\
&& = - |g|^2 \Lambda^2 \int_p \frac{K(p) (1 - K(p))}{p^2} +
\mathrm{O}\left(|g|^2 |m|^2\right)\nn\\
&& = - |g|^2 \Lambda^2 \frac{1}{2} \int_p \frac{\Delta (p) - 2 \Delta
  (p) (1 - K(p))}{p^2} + \mathrm{O}\left(|g|^2 |m|^2\right)
\end{eqnarray}
reproducing $a_4^{(1)}$ of (\ref{a49}).

We must conclude that quadratic divergences generally exist in the
Wilson action of the Wess-Zumino model.  This is analogous to the
presence of a quadratically divergent gauge boson mass term in the
Wilson actions of gauge theories.  (For an explicit calculation in
QED, see \cite{qed}.)  We expect that any realization of supersymmetry
on a lattice (if it exists) is analogous to the choice $K_b \ne K_f$,
and that fine-tuning of order $\mu^2/\Lambda^2$ will be inevitable.

Second on holomorphy.  In \cite{seiberg} the non-renormalization of
the F-term of the Wilson action in superfields has been explained as a
consequence of holomorphy, i.e., the F-term depends either only on
$\phi, \chi_R, F$, $g, m$, or only on $\bar{\phi}, \chi_L, \bar{F}$,
$\bar{g}, \bar{m}$. This concept of holomorphy has been introduced
with help of the $\alpha$, $\beta$ charges and of the idea of local
couplings that is inspired by string theory. The non-renormalization
theorem has been also derived in \cite{fk} based on the observation
that the chiral vertices can be written as multiple supersymmetry
variations.  The approach of \cite{fk} has been further generalized to
the component field formalism with on-shell supersymmetry
\cite{ulker}.  Thus, it is possible to formulate holomorphy in terms
of component fields. We have not, however, studied how to incorporate
holomorphy in the ERG formalism.  Nevertheless we can expect that
holomorphy corresponds to some constraints on the free parameters
$z_{1,2,5,9} (0)$ of the theory.  Examining the 1-loop results of
sect.~\ref{oneloop}, we notice $z_{2,5,9} (\Lambda/\mu)$ have no
dependence on $\Lambda/\mu$.  We then speculate that holomorphy is
equivalent to
\begin{equation}
z_2 (\ln \Lambda/\mu) = z_5 (\ln \Lambda/\mu) = z_9 (\ln \Lambda/\mu) = 0
\end{equation}
under the following choice of the parameters
\begin{equation}
z_2 (0) = z_5 (0) = z_9 (0) = 0
\end{equation}
The verification is left for a future study.

\section{Conclusions}

In this paper we have constructed a Wilson action of the Wess-Zumino
model using ERG perturbatively: solving the exact renormalization
group differential equation and imposing supersymmetry.  The action is
built out of scalar and spinor fields without auxiliary fields, and
satisfies the invariance (\ref{master}) under the supersymmetry
transformation (\ref{susy}).  In order to prove the consistency of the
construction, we have resorted to the BRST formalism, by introducing
antifields that generate the transformation.  But the antifields are
necessary only for the proof, and once we know the construction works,
we can discard the antifields entirely.  Hence, the Wilson action is a
functional only of scalar and spinor fields.

The Wilson action, thus constructed, has four arbitrary parameters:
\begin{enumerate}
\item the mass parameter $m$
\item the coupling parameter $g$
\item the common normalization of scalar and spinor fields --- the
    relative normalization is fixed by our choice of supersymmetry
    transformation 
\item the constant shift of $\phi, \bar{\phi}$ proportional to
    $\bar{g} m, g \bar{m}$ --- physics does not change, but the action
    changes its appearance
\end{enumerate}
In the formulation without auxiliary fields, each parameter gets its
own beta function that gives the dependence of the parameter on the
renormalization scale $\mu$.  How the beta functions arise in the
context of ERG has been discussed in ref.~\cite{erg}.  The derivation
of the beta functions for the Wess-Zumino model is left for a future
study.  Since our construction of the supersymmetric Wilson action
depends only on its supersymmetry, we expect that the ERG method can
be applied straightforwardly to supersymmetric Yang-Mills theories and
theories with extended supersymmetry; this is also left for future
studies.

\section*{Acknowledgements}

H.S. wishes to thank the members of the Feza G\"ursey Institute,
Istanbul, for their hospitality in September 2007.  The collaboration
for this work started during the visit.  We thank Prof.~M.~Sakamoto
for a discussion on regularization of supersymmetry, and also thank
Dr.~L.~Akant for his participation in an early stage of this work.

\appendix

\section{Notation on spinors}

Throughout this paper we work on the four dimensional euclidean
space. In this appendix we summarize the basic properties of
spinors, emphasizing what is relevant to this paper.

Denoting the orthogonal coordinates by $x_\mu\,(\mu=1,2,3,4)$, we
define
\begin{equation}
\mathbf{x} \equiv i x_\mu \sigma_\mu
\end{equation}
where $\sigma_\mu$ are defined by the Pauli matrices and the 2-by-2 unit
matrix as
\begin{equation}
\sigma_\mu \equiv (\vec{\sigma}, -i \mathbf{1}_2)
\end{equation}
We then find
\begin{equation}
\det \mathbf{x} = x_\mu x_\mu = x^2
\end{equation}
An arbitrary rotation can be given as
\begin{equation}
\mathbf{x} \longrightarrow \mathbf{x}' = L \mathbf{x}
R^{-1} \label{rotation} 
\end{equation}
where $L, R$ are arbitrary SU(2) matrices, and hence $\det \mathbf{x}$
is invariant under any rotation.

Under the rotation (\ref{rotation}) a right-hand two-component spinor
field $\chi_R (x)$ transforms as
\begin{equation}
\chi_R (x) \longrightarrow \chi'_R (x') = R \chi_R (x)
\end{equation}
Similarly, a left-hand two-component spinor field $\chi_L (x)$
transforms as
\begin{equation}
\chi_L (x) \longrightarrow \chi'_L (x') = L \chi_L (x)
\end{equation}
Denoting the transpose of a spinor (either right or left) by
\begin{equation}
\bar{\chi} \equiv \chi^T \sigma_y = (i \chi_2, - i \chi_1)
\end{equation}
we can construct scalars:
\begin{equation}
\bar{\chi}_R \chi'_R,\quad \bar{\chi}_L \chi'_L
\end{equation}
We obtain
\begin{equation}
\bar{\chi}_R \chi'_R = \pm \bar{\chi}'_R \chi_R,\quad
\bar{\chi}_L \chi'_L = \pm \bar{\chi}'_L \chi_L
\end{equation}
depending on whether the two spinors are mutually anticommuting (plus)
or commuting (minus).

We call the hermitian conjugate of $\sigma_\mu$ by
\begin{equation}
\bar{\sigma}_\mu \equiv \sigma_\mu^\dagger = - \sigma_y \sigma_\mu^T
\sigma_y
\end{equation}
Using $\sigma_\mu$ or $\bar{\sigma}_\mu$, we can construct a vector:
\begin{equation}
\bar{\chi}_L \sigma_\mu \chi_R = \mp \bar{\chi}_R \bar{\sigma}_\mu  \chi_L
\end{equation}
Here, the sign is minus for mutually anticommuting
spinors, and plus for mutually commuting spinors.

For mutually commuting spinors (either R or L),the following
identity holds:
\begin{equation}
\chi (\bar{\chi}' \chi'') + \chi' (\bar{\chi}'' \chi) + \chi''
(\bar{\chi} \chi') = 0
\end{equation}
We have used this often in our calculations.

\section{Classical BRST invariance}

In this appendix we verify the classical BRST invariance of the
classical action, defined by
\begin{eqnarray}
S_{cl} &\equiv& - \int d^4 x \left[ \bar{\chi}_L \sigma_\mu \partial_\mu
    \chi_R + \partial_\mu \bar{\phi} \partial_\mu \phi + \bar{F}
    (\bar{\phi}) F (\phi)\right.\nn\\
&&\qquad\qquad \left. + F' (\phi) \frac{1}{2} \bar{\chi}_R \chi_R
    + \bar{F}' (\bar{\phi}) \frac{1}{2} \bar{\chi}_L \chi_L \right]
\end{eqnarray}
where $F (\phi)$ is an arbitrary function of $\phi$, and $\bar{F}$ its
complex conjugate.  $F'$ is the derivative of $F$ with respect to
$\phi$.  We introduce a supersymmetry transformation as the following
BRST transformation:
\begin{equation}
\lb\begin{array}{c@{~=~}l}
\delta \phi & \bar{\xi}_R \chi_R + \eta_\mu \partial_\mu \phi\\
\delta \bar{\phi} & \bar{\xi}_L \chi_L + \eta_\mu \partial_\mu
\bar{\phi}\\
\delta \chi_R & \bar{\sigma}_\mu \xi_L \partial_\mu \phi - \bar{F}
(\bar{\phi}) \xi_R + \eta_\mu \partial_\mu \chi_R\\
\delta \chi_L & \sigma_\mu \xi_R \partial_\mu \bar{\phi} - F(\phi)
\xi_L + \eta_\mu \partial_\mu \chi_L
\end{array}\right.\label{classicalsusy}
\end{equation}
where $\xi_{R,L}$ are constant spinors, and $\eta_\mu$ a constant
vector.  We take $\xi_{R,L}$ commuting, and $\eta_\mu$ anticommuting.
It is straightforward to check the BRST invariance of $S_{cl}$:
\begin{equation}
\delta S_{cl} \equiv \int d^4 x \left[ \frac{\delta S_{cl}}{\delta \phi}
    \delta \phi + 
    \frac{\delta S_{cl}}{\delta \bar{\phi}} \delta \bar{\phi} + S_{cl}
    \Rd{\chi_R} \delta \chi_R + S_{cl} \Rd{\chi_L} \delta \chi_L
\right] = 0
\end{equation}

The BRST transformation (\ref{classicalsusy}) is not nilpotent.
Defining the transformation of $\xi_{R,L}$ and $\eta_\mu$ by
\begin{equation}
\delta \xi_{R,L} = 0,\quad \delta \eta_\mu = - \bar{\xi}_R
\bar{\sigma}_\mu \xi_L = - \bar{\xi}_L \sigma_\mu \xi_R
\end{equation}
we obtain
\begin{equation}
\lb\begin{array}{c@{~=~}l}
\delta^2 \phi & 0\\
\delta^2 \bar{\phi} & 0\\
\delta^2 \chi_R & \xi_R \cdot \bar{\xi}_L \Ld{\bar{\chi}_L} S_{cl}\\
\delta^2 \chi_L & \xi_L \cdot \bar{\xi}_R \Ld{\bar{\chi}_R} S_{cl}
\end{array}\right.
\end{equation}
Hence, to make $\delta$ nilpotent, we must invoke the equations of
motion:
\begin{equation}
\Ld{\bar{\chi}_L} S_{cl} = 0,\quad \Ld{\bar{\chi}_R} S_{cl} = 0
\end{equation}
In other words, the algebra of the classical supersymmetry
transformation is closed only on the mass shell.

The standard way to elevate the on-shell algebra to the off-shell
algebra is to introduce auxiliary fields.  Here, we take a different
route, however.  We introduce antifields: fermionic scalars $\phi^*,
\bar{\phi}^*$ and bosonic spinors $\chi_R^*, \chi_L^*$.  We first
couple them linearly to the supersymmetry transformation:
\begin{equation}
S_{cl,1} \equiv S_{cl} + \int d^4 x \left[ \phi^* \delta \phi +
    \bar{\phi}^* \delta \bar{\phi} + \bar{\chi}_R^* \delta \chi_R +
    \bar{\chi}_L^* \delta \chi_L \right]
\end{equation}
so that
\begin{equation}
\lb\begin{array}{c@{~=~}l@{\quad}c@{~=~}l}
\Ld{\phi^*} S_{cl,1} & \delta \phi,& \Ld{\bar{\phi}^*} S_{cl,1} &
\delta \bar{\phi}\\
\Ld{\bar{\chi}_R^*} S_{cl,1} & \delta \chi_R,& \Ld{\bar{\chi}_L^*}
  S_{cl,1} & \delta \chi_L
\end{array}\right.
\end{equation}
We now transform $S_{cl,1}$:
\begin{eqnarray}
    &&\delta S_{cl,1} \equiv \int d^4 x \left[ \frac{\delta
          S_{cl,1}}{\delta \phi} \Ld{\phi^*} 
        S_{cl,1} + \frac{\delta S_{cl,1}}{\delta \bar{\phi}}
        \Ld{\bar{\phi}^*} S_{cl,1} \right.\nn\\
    &&\quad \left.- S_{cl,1} \Rd{\chi_R} \Ld{\chi_R^*}
        S_{cl,1} - S_{cl,1} \Rd{\chi_L} \Ld{\chi_L^*} S_{cl,1} \right] -
    \bar{\xi}_L \sigma_\mu \xi_R \frac{\overrightarrow{\partial}}{\partial
      \eta_\mu} S_{cl,1}
\end{eqnarray}
This can be calculated further as
\begin{eqnarray}
\delta S_{cl,1} &=& \int \left[ - \phi^* \delta^2 \phi - \bar{\phi}^* \delta^2
        \bar{\phi} + \bar{\chi}_R^* \delta^2 \chi_R + \bar{\chi}_L^*
        \delta^2 \chi_L \right]\nn\\
    &=& \int \left[ \bar{\chi}_R^* \xi_R \cdot \bar{\xi}_L
        \Ld{\bar{\chi}_L} S_{cl} + \bar{\chi}_L^* \xi_L \cdot \bar{\xi}_R
        \Ld{\bar{\chi}_R} S_{cl} \right]\nn\\
    &=& \int \left[ \bar{\chi}_R^* \xi_R \cdot \bar{\xi}_L
        \Ld{\bar{\chi}_L} S_{cl,1} + \bar{\chi}_L^* \xi_L \cdot \bar{\xi}_R
        \Ld{\bar{\chi}_R} S_{cl,1} \right]
\end{eqnarray}
To cancel this, we modify the action by adding a term quadratic in
antifields: 
\begin{equation}
\bar{S}_{cl} \equiv S_{cl,1} - \int d^4 x \,\bar{\chi}_R^* \xi_R \cdot
\bar{\chi}_L^* \xi_L
\end{equation}
such that
\begin{equation}
\lb\begin{array}{c@{~=~}l}
\Ld{\chi_R^*} \bar{S}_{cl} & \Ld{\chi_R^*} S_{cl,1} - \xi_R \cdot
\bar{\chi}_L^* \xi_L\\
\Ld{\chi_L^*} \bar{S}_{cl} & \Ld{\chi_L^*} S_{cl,1} - \xi_L \cdot
\bar{\chi}_R^* \xi_R
\end{array}\right.
\end{equation}
This satisfies the classical BRST invariance (or Zinn-Justin
equation):
\begin{eqnarray}
&&\int d^4 x \left[ \frac{\delta \bar{S}_{cl}}{\delta \phi} \Ld{\phi^*}
    \bar{S}_{cl} + \frac{\delta \bar{S}_{cl}}{\delta \bar{\phi}}
    \Ld{\bar{\phi}^*} \bar{S}_{cl}\right.\nn\\
&&\quad \left. - \bar{S}_{cl} \Rd{\chi_R}
    \Ld{\chi_R^*} \bar{S}_{cl}  - \bar{S}_{cl} \Rd{\chi_L}
    \Ld{\chi_L^*} \bar{S}_{cl}  \right] - \bar{\xi}_L \sigma_\mu \xi_R 
\frac{\overrightarrow{\partial}}{\partial
  \eta_\mu} \bar{S}_{cl} = 0
\end{eqnarray}

\section{Antifield dependence of $\tilde{\Sigma}_l$}

In this appendix, we wish to show
\begin{enumerate}
\item $\tilde{\Sigma}^{(l > 0)}$ is independent of $\eta_\mu, \phi^*,
    \bar{\phi}^*$.
\item The dependence of $\tilde{\Sigma}^{(l > 0)}$ on
    $\chi_R^*, \chi_L^*$ comes only through $j_R, j_L$.
\end{enumerate}

We expand $\tilde{S}, \tilde{\Sigma}$ in the number of loops:
\begin{equation}
\tilde{S} = \sum_{l=0}^\infty \tilde{S}^{(l)},\quad
\tilde{\Sigma} = \sum_{l=1}^\infty \tilde{\Sigma}^{(l)}
\end{equation}
Only $\tilde{S}^{(0)}$ has dependence on $\eta_\mu, \phi^*,
\bar{\phi}^*$; $\tilde{S}^{(l)}\,(l > 0)$ depends only on $j_R, j_L$
and regular fields $\phi,\bar{\phi},\chi_R, \chi_L$.  Therefore, we
obtain
\begin{eqnarray}
\tilde{\Sigma} &=& \int_p \K{p} \Bigg[ \Ld{\phi^* (-p)} \tilde{S}^{(0)}
\cdot \frac{\delta \tilde{S}}{\delta \phi (p)} + \Ld{\bar{\phi}^* (-p)}
\tilde{S}^{(0)} \cdot \frac{\delta \tilde{S}}{\delta \bar{\phi} (p)}\nn\\
&&\, - \tilde{S} \Rd{\chi_R (p)} \cdot \Ld{\bar{\chi}_R^* (-p)}
\tilde{S} - \Tr \Ld{\bar{\chi}_R^* (-p)} \tilde{S} \Rd{\chi_R (p)}\nn\\
&&\, - \tilde{S} \Rd{\chi_L (p)} \cdot \Ld{\bar{\chi}_L^* (-p)}
\tilde{S} - \Tr \Ld{\bar{\chi}_L^* (-p)} \tilde{S} \Rd{\chi_L (p)}
\quad \Bigg]\nn\\
&&\quad - \bar{\xi}_L \sigma_\mu \xi_R \cdot
\frac{\overrightarrow{\partial}}{\partial \eta_\mu} \tilde{S}^{(0)}
\end{eqnarray}
Substituting the loop expansions of $\tilde{S}$ into the above, we
obtain
\begin{eqnarray}
&&\tilde{\Sigma}^{(l > 0)} = \int \K{p} \left[ \Ld{\phi^* (-p)}
\tilde{S}^{(0)} \cdot \frac{\delta \tilde{S}^{(l)}}{\delta \phi (p)} +
\Ld{\bar{\phi}^* (-p)} \tilde{S}^{(0)} \cdot \frac{\delta
  \tilde{S}^{(l)}}{\delta \bar{\phi} (p)}\right.\nn\\
&&\,- \sum_{l'=0}^l \tilde{S}^{(l-l')} \Rd{\chi_R (p)} \Ld{\bar{\chi}_R^*
  (-p)} \tilde{S}^{(l')} - \Tr \Ld{\bar{\chi}_R^* (-p)} \tilde{S}^{(l-1)}
\Rd{\chi_R (p)}\nn\\
&&\, \left. 
- \sum_{l'=0}^l \tilde{S}^{(l-l')} \Rd{\chi_L (p)} \Ld{\bar{\chi}_L^*
  (-p)} \tilde{S}^{(l')} - \Tr \Ld{\bar{\chi}_L^* (-p)} \tilde{S}^{(l-1)}
\Rd{\chi_L (p)} \right]
\end{eqnarray}

Now, recall (\ref{Stilde}):
\begin{eqnarray}
    \tilde{S} &\equiv& \Sf [\phi,\bar{\phi},\chi_R,\chi_L] + \hSi [\phi,
\bar{\phi}, \chi_R, \chi_L; j_R, j_L] \nn\\ 
    && + \int_p \frac{1}{\K{p}} \Big[ \quad\phi^* (-p) \left\lbrace i p_\mu
        \eta_\mu \phi (p) + \bar{\xi}_R \chi_R (p) \right\rbrace \nn\\
    &&\quad +
    \bar{\phi}^* (-p) \left\lbrace i p_\mu \eta_\mu \bar{\phi} (p) +
        \bar{\xi}_L \chi_L (p) \right\rbrace\nn\\
    && \quad + \bar{\chi}_R^* (-p) \left\lbrace i p_\mu \eta_\mu \chi_R (p) +
        i p \cdot \bar{\sigma} \xi_L \phi (p) - \bar{m} \xi_R \bar{\phi}
        (p) \right\rbrace\nn\\
    && \quad + \bar{\chi}_L^* (-p) \left\lbrace i p_\mu \eta_\mu \chi_L (p) +
        i p \cdot \sigma \xi_R \bar{\phi} (p) - m \xi_L \phi (p)
    \right\rbrace\nn\\
    &&\quad - j_R (-p) j_L (p) \quad\Big]
\end{eqnarray}
Since $\tilde{S}^{(l)}$ for $l > 0$ is the $l$-loop part of $\hSi$, we
obtain
\begin{equation}
\Ld{\bar{\chi}_R^*} \tilde{S}^{(l)} = \xi_R \frac{\delta
  \tilde{S}^{(l)}}{\delta j_R},\quad
\Ld{\bar{\chi}_L^*} \tilde{S}^{(l)} = \xi_L \frac{\delta
  \tilde{S}^{(l)}}{\delta j_L}
\end{equation}
Therefore, we get
\begin{eqnarray}
&&\tilde{\Sigma}^{(l)} = \eta_\mu \int_p i p_\mu \Bigg[ \,
\phi (p) \frac{\delta \tilde{S}^{(l)}}{\delta \phi (p)}
+ \bar{\phi} (p) \frac{\delta \tilde{S}^{(l)}}{\delta \bar{\phi} (p)}\nn\\
&&\qquad + \tilde{S}^{(l)} \Rd{\chi_R (p)} \chi_R (p) +
\tilde{S}^{(l)} \Rd{\chi_L (p)} \chi_L (p) \nn\\
&&\qquad - \bar{\chi}_R^* (-p) \Ld{\bar{\chi}_R^* (-p)} \tilde{S}^{(l)}
- \bar{\chi}_L^* (-p) \Ld{\bar{\chi}_L^* (-p)} \tilde{S}^{(l)}\,\Bigg]\nn\\
&& \,- \int_p \left[ \phi^* (-p) \bar{\xi}_R \xi_R \frac{\delta
      \tilde{S}^{(l)}}{\delta j_R (-p)} + \bar{\phi}^* (-p)
    \bar{\xi}_L \xi_L \frac{\delta \tilde{S}^{(l)}}{\delta
      j_L (-p)} \right]\nn\\
&& \,+ \int_p \K{p} \left[\,
\frac{1}{\K{p}} \lb \bar{\xi}_R \chi_R (p) \frac{\delta
  \tilde{S}^{(l)}}{\delta \phi (p)} + \bar{\xi}_L \chi_L (p) \frac{\delta
  \tilde{S}^{(l)}}{\delta \bar{\phi} (p)} \rb\right.\nn\\
&&\quad - \sum_{l'=1}^{l-1} \tilde{S}^{(l-l')} \Rd{\chi_R (p)} \xi_R
\frac{\delta \tilde{S}^{(l')}}{\delta j_R (-p)} \nn\\
&&\quad - \frac{1}{\K{p}} \tilde{S}^{(l)} \Rd{\chi_R (p)} \left( i p \cdot
  \bar{\sigma} \xi_L \phi (p) - \bar{m} \xi_R \bar{\phi} (p) + \xi_R
  \frac{\delta \hSi^{(0)}}{\delta j_R (-p)} \right)\nn\\
&&\quad - \left( \Sf + \hSi^{(0)} \right)
 \Rd{\chi_R (p)} \xi_R \frac{\delta \tilde{S}^{(l)}}{\delta j_R
  (-p)} -  \frac{\delta \tilde{S}^{(l)}}{\delta j_R (-p)} \Rd{\chi_R (p)}
\xi_R\nn\\
&&\quad - \sum_{l'=1}^{l-1} \tilde{S}^{(l-l')} \Rd{\chi_L (p)} \xi_L
\frac{\delta \tilde{S}^{(l')}}{\delta j_L (-p)} \nn\\
&&\quad - \frac{1}{\K{p}} \tilde{S}^{(l)} \Rd{\chi_L (p)} \left( i p \cdot
  \sigma \xi_R \bar{\phi} (p) - m \xi_L \phi (p) + \xi_L
  \frac{\delta \hSi^{(0)}}{\delta j_L (-p)} \right)\nn\\
&&\quad \left.- \left( \Sf + \hSi^{(0)} \right) \Rd{\chi_L (p)} \xi_L
\frac{\delta \tilde{S}^{(l)}}{\delta j_L 
  (-p)} -  \frac{\delta \tilde{S}^{(l)}}{\delta j_L (-p)} \Rd{\chi_L (p)}
\xi_L\,\right]
\end{eqnarray}
The first part that is proportional to $\eta_\mu$ vanishes due to the
translation invariance of $\tilde{S}^{(l)}$.  Due to  the Bose
statistics of $\xi_R, \xi_L$, we obtain
\begin{equation}
\bar{\xi}_R \xi_R = \bar{\xi}_L \xi_L = 0
\end{equation}
and this makes the second part zero.  What is left is built out of
$\Sf + \hSi^{(0)}$ and $\tilde{S}^{(1)}, \cdots, \tilde{S}^{(l)}$.
Hence, $\tilde{\Sigma}^{(l)}$ satisfies the two properties stated at
the beginning of this appendix.

\end{document}